\documentclass[useAMS]{mn2e}
\usepackage{times}
\usepackage{aas_symbols}
\usepackage{rotating}
\usepackage{color}
\long\def\symbolfootnote[#1]#2{\begingroup%
\def\thefootnote{\fnsymbol{footnote}}\footnote[#1]{#2}\endgroup}
\def\G{\Gamma}
\def\g{\gamma}
\def\D{\Delta}
\def\eps{\epsilon}
\def\b{\beta}
\def\th{\theta}
\def\et{\eta}
\def\del{\delta}

\begin{document}

\title[Late Jet, SNa Ejecta and Cocoon in GRBs]
{The late jet in gamma-ray bursts and its interactions with a supernova ejecta and a cocoon}

\author[Shen, Kumar \& Piran]{Rongfeng Shen$^{1}$\thanks{E-mail: rfshen@astro.as.utexas.edu (RS); pk@astro.as.utexas.edu (PK); tsvi@phys.huji.ac.il (TP)}, Pawan Kumar$^{1}$\footnotemark[1] and Tsvi Piran$^{2}$\footnotemark[1]\\
$^{1}$Department of Astronomy, University of Texas, Austin, TX 78712, USA\\
$^{2}$Racah Institute of Physics, Hebrew University of Jerusalem, Jerusalem 91904, Israel}

\date{}
\maketitle

\begin{abstract}

Late X-ray flares observed in X-ray afterglows of
gamma-ray bursts (GRBs) suggest
late central engine activities at a few minuets to hours after the burst. A few
unambiguously confirmed
cases of supernova associations with nearby long GRBs imply that an
accompanying supernova-like component might be a common feature in all
long GRB events. These motivate us to study the interactions of a late
jet, responsible for a x-ray flare, with various components
in a stellar explosion, responsible for a GRB. These components include a supernova shell-like ejecta,
and a cocoon that was produced when the main jet producing the
GRB itself was propagating through the progenitor star. We find that the interaction between the late jet and the supernova
ejecta may produce a luminous (up to $10^{49}$ erg s$^{-1}$) thermal X-ray transient lasting for $\sim 10$ s
The interaction between the late jet and the cocoon produces
synchrotron-self absorbed non-thermal emission, with the observed peak
X-ray flux density from 0.001 $\mu$Jy to 1 mJy at 1 keV and a peak optical
flux density from 0.01 $\mu$Jy to 0.1 Jy (for a redshift $z= 2$). The light curve due to the
late jet - cocoon interaction has very small pulse-width-to-time ratio,
$\D t/t \approx 0.01 - 0.5$, where $t$ is the pulse peak time since the
burst trigger. Identifying these features in current and future
observations would open a new frontier in the study of GRB progenitor
stars.
\end{abstract}



\section{Introduction}

Long duration gamma-ray bursts (GRB) -- those lasting more than 2 s -- are
thought to be produced by a relativistic outflow
(or jet) with a kinetic energy $\sim 10^{51} - 10^{52}$ erg (beaming
effect corrected) when a massive star collapses at the end of its nuclear
burning life cycle  (see Piran 1999, 2005; M\'{e}sz\'{a}ros
2002 for reviews). The massive star origin of GRBs are supported by two different
observations: (i) GRBs are found to be in actively star forming galaxies
(e.g., Christensen et al. 2004, Castro Cer\'on et al. 2006) or in
star (especially massive ones) forming regions of the host galaxies
(e.g., Paczy\'{n}ski 1998; Bloom, Kulkarni \& Djorgovski 2002; Fruchter et al. 2006);
(ii) for a subset of nearly a dozen of
GRBs, X-ray-rich GRBs and X-ray flashes, Supernova (SNa) features
-- both temporally and spacially associated with the bursts
(Woosley \& Bloom 2006 for a review) -- were detected.
For four of those: GRB 980425 (e.g., Galama et al. 1998), 030329
(e.g., Hjorth et al. 2003), 021211 (Della Valle et al. 2003) and 031203
(e.g., Malesani et al., 2004), the physically associated SNe were
not only photometrically but also spectroscopically confirmed.
The others of the subset show a late-time ($\sim$ 10 days) SNa-like
``bump'' in the optical afterglows, with a simultaneous strong
color evolution, e.g., in GRB 980326 (Bloom et al. 1999) and 011121
(Bloom et al. 2002), consistent with the hypothesis of an underlying SNa.

The {\it Swift} satellite has recently unveiled a ``canonical'' behavior pattern in about two-thirds of GRBs' early X-ray afterglows: a rapid decline phase lasting for $\sim 10^2$ s is followed by a shallow decay phase lasting $\sim 10^3 - 10^4$ s,
then by a ``normal'' power-law decay phase and finally by a possible jet break (Nousek et al. 2006; O'Brien et al. 2006). In addition, X-ray flares are found in about 50\% of all {\it Swift} bursts; they have been discovered in all of the above
phases (Burrows et al. 2005, 2007; Chincarini et al. 2007). Even long before {\it Swift}, a late X-ray flare was detected by BeppoSAX for GRB 970508 (Piro et al. 1998).

In this work we investigate a scenario in which a late jet -- responsible for producing the X-ray flares and possibly the shallow decay phase -- interacts with other components of a long-GRB stellar explosion. These components include a SNa ejecta, if a SNa explosion is accompanying the GRB event, and a cocoon created by the passage of the main GRB jet through the star (Ramirez-Ruiz et al. 2002, Matzner 2003, Zhang et al. 2004).

In Section \ref{sec:mul-com} we argue for the existence of the late jet and multiple components of a GRB explosion, and provide physical motivations of this work. We investigate the interactions of the late jet with the SNa ejecta in Section \ref{sec:jet-ejecta} and with the cocoon in Section \ref{sec:jet-cocoon} and calculate their associated emissions. The predicted emissions and their detection prospects are confronted with current observational data in Section \ref{sec:obs}. The summary and implications are given in Section \ref{sec:summary}.


\section{A multi-component GRB stellar explosion}\label{sec:mul-com}

Two recently discovered GRB features point to the emergence of a late
outflow (jet) after the main $\g$-ray producing outflow has died.
The first is the X-ray flares observed at a few $\times 10^2 - 10^3$ s
(as late as 10$^4-10^5$ s in some cases) after the prompt burst,
with a fluence typically about one tenth of the fluence of the
prompt $\g$-rays; in one case, GRB 050502B, this ratio is
$\sim$ 1 (Falcone et al. 2007). Lazzati, Perna \& Begelman (2008) found that the
mean flux of a flare declines with its occurrence time as $\sim t^{-1.5}$.
These flares are characterized by a large flux
increase and by a very steep rise and decay. Typical increase of the flux ranges from a factor
of order unity to 10 and in some rare cases even a few hundred.
The decay after the flare peak is as steep as
$\propto t^{-4}$, much steeper than the underlying afterglow decay
($\propto t^{-1}$). The pulse width to the peak time ratio $\D t/t$
is much smaller than unity, typically $\sim 0.3$ (Burrows et al. 2005, 2007;
Chincarini et al. 2007).

The late flares cannot be due to external-origin
mechanisms such as, a density clump in the circumburst medium,
or the energy injection into the afterglow blastwave by the trailing
slower shells, because the decay slope after the rebrightening from
these mechanisms always follows the standard afterglow model, and
$\D t/ t \sim 1$ is always expected (Nakar, Piran \& Granot 2003;
Nakar \& Granot 2007; Lazzati \& Perna 2007).
They are also unlikely to arise from
late collisions between two slow moving shells ejected at same
time as the main $\g$-ray producing shells, since the resultant internal
shock is too weak to give rise to the significant emission observed in
flares (Zhang 2006). The most likely possibility for X-ray flares is the late activity
of the central engine (e.g., Fan \& Wei 2005).
Such a late activity was proposed already in 1998 as an alternative
origin of GRB afterglow (Katz \& Piran 1998, Katz, Piran \& Sari 1998).

The scenario involving the late engine activity can easily satisfy
the constraint that $\D t /t \ll 1$.
Also in this scenario the late flare is physically
separate from the ``background'' afterglow, so the large amplitude
increase of the flux superposed on the decaying afterglow can be
naturally explained.

The second feature favoring the existence of a late jet is the shallowly
decaying component in the overall X-ray light curve.
This component starts at a few $\times 10^2 - 10^3$ s and ends at
$10^4-10^5$ s during which the flux decays slower than
expected for standard decelerating blast-wave afterglow.
One straightforward interpretation is that the shallowness is due to an
long-lasting energy injection from a late outflow that catches up with the decelerated main outflow
(Cohen \& Piran 1999; Zhang \& M\'{e}sz\'{a}ros 2001;
Yu \& Dai 2007), though there are other possible interpretations for this
feature (e.g., Granot \& Kumar 2006; Granot, K\"{o}nigl \& Piran 2006; Fan \& Piran 2006,
Panaitescu et al. 2006; Kobayashi \& Zhang 2007; Uhm \& Beloborodov 2007;
Genet et al. 2007).

In those cases where we don't see a SNa spectroscopically or photometrically
in the optical band, we can still hope to
explore the existence and properties of the SNa accompanying the GRB
by looking at the interaction of a late jet with the SNa remnant and the
emission from it.
It is likely that every long-duration GRB has a
SNa accompanying it but some extrinsic and/or intrinsic bias might have
hindered the optical detection of the SNa component(Woosley \& Bloom 2006).
A late jet provides a chance to test this picture in the cases where
the ordinary SNa features are not easy to detect.

\begin{figure*}
\centerline{\includegraphics[width=12cm,angle=0]{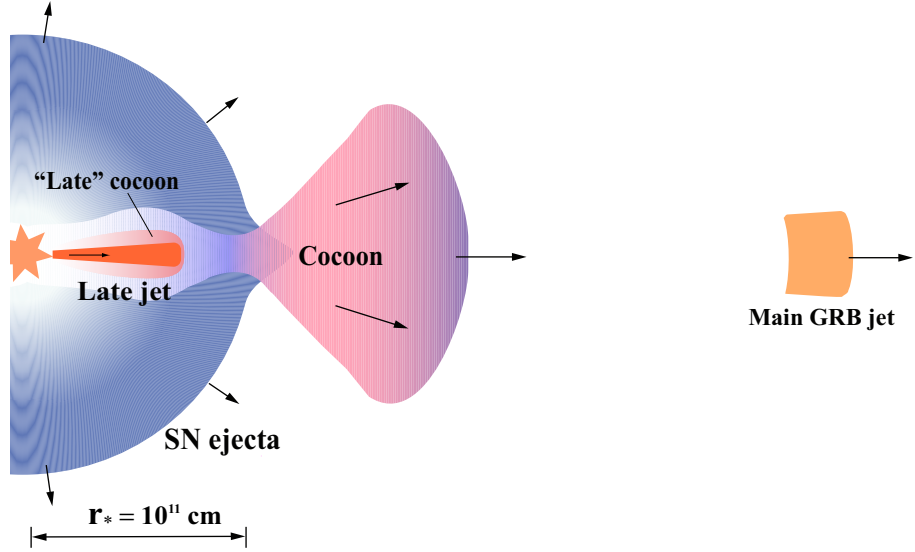}}
\caption{Schematic illustration of multiple components in
a long GRB and an accompanying supernova (SNa). An initial, highly relativistic jet,
as shown in the right end of the illustration, is responsible for the
prompt GRB. A cocoon, that was inflated by the initial GRB jet as the jet was
punching through the envelope of the progenitor star, has broken out of
the stellar surface at the same time as the main jet, and
has accelerated to mildly relativistic speed. A nearly spherically symmetric
sub-relativistic ejecta is responsible for the SNa. A late jet responsible
for the late X-ray flares in GRBs is launched from the central source
and will catch up with the above components. The ``late'' cocoon is produced
when the late jet crosses the SNa ejecta. The distances are not to scale.}
\label{fig:jet-illus}
\end{figure*}

Given the evidences for the existence of a late jet and the physical
association of SNe and GRBs, one expects a number of different interactions
between the following four components in a GRB event (see Fig.
\ref{fig:jet-illus} for an illustration): (i) A
highly relativistic ($\G \sim 300 - 1000$) narrow jet along the rotation
axis with an opening angle $\sim$ 0.1 radian that produces the main
GRB event; (ii) A nearly spherically symmetric SNa ejecta moving with
speed $\sim 10^4$ km s$^{-1}$; (iii) A cocoon fireball created by
the passage of the main GRB jet through the
star;
(iv) A late relativistic jet responsible for the late X-ray flares after
the end of the main GRB. See Woosley \& Heger (2006) for a detailed
version of the multi-component GRB scenario. Also see Wheeler et al.
(2000) for a similar but more detailed model which relates various
energetic phenomena such as SNa, GRB and magnetar in a single stellar
explosion.

In this work we investigate the interactions of a late jet which might have a
similar Lorentz factor and opening angle to those of the main GRB
jet, with the expanding SNa ejecta and the cocoon.
We assume that the SNa ejecta and the main GRB jet are launched
from the central source at about the same time
\footnote{The observational constraint is that GRB and SNa occur within
$\sim$ 1 day (e.g., Woosley \& Bloom 2006). SNa explosion theories estimate
that the SNa shock breaks out at a few tens of seconds after the core rebounce
(e.g., Janka et al. 2007 and references therein); this time scale
is small compared to the delay of the late jet ($t_F \ge 10^2$ s).}. Wheeler et al. (2002)
described a similar stellar collapsing scenario where a delayed
relativistic jet from the eventually formed black hole catches up and collides
with an earlier proto-pulsar toroidal field generated mildly relativistic jet
as an origin of the $\g$-ray burst.
Ghisellini et al. (2007) firstly considered the
collision between a jetted fireball from an intermittent GRB central engine and a stationary
cocoon as an alternative to the standard internal shock scenario, but with higher efficiency.
Here we are not attempting to explain how a relativistic jet is formed and a $\g$-ray
burst is produced from the relativistic jet or how an accompanying SNa
is generated, rather we are trying to constrain the physical properties of
the late jet, SNa ejecta and the cocoon by calculating the emissions
from their interactions, and to verify the general picture of GRB-SNa connection.

We will use the following fiducial values for a variety of model
parameters. For the late jet, a total energy $E_j \sim 10^{51}$
erg, opening angle $\theta_j \sim$ 0.1 radian, LF
$\G_j \sim$ 100, a delay respective to the launching of the main GRB jet,
cocoon and SNa ejecta (in our picture the three are launched at more or
less the same time) $t_F \sim 10^2$ s, and a duration $t_{dur} \sim 10^2$ s;
for the SNa ejecta, we use an isotropic-equivalent mass $M_{SN} \sim$ 10
$M_{\odot}$ and speed $V_{SN} \sim 10^9$ cm s$^{-1}$; for the cocoon,
we use an energy $E_{c} \sim 10^{51}$ erg, a terminal LF $\G_c \sim$
10 and an opening angle $\theta_c \sim$ 0.6 radian.
Nevertheless an appropriate range of numerical values is
assigned to each parameter in the real calculation (e.g., $E_j$ could
be two orders of magnitude larger or smaller than the fiducial one, cocoon
speed could be sub-relativistic and $t_F$ could be as large as $10^4-10^5$
s).

\begin{figure}
\includegraphics[height=6cm,angle=0]{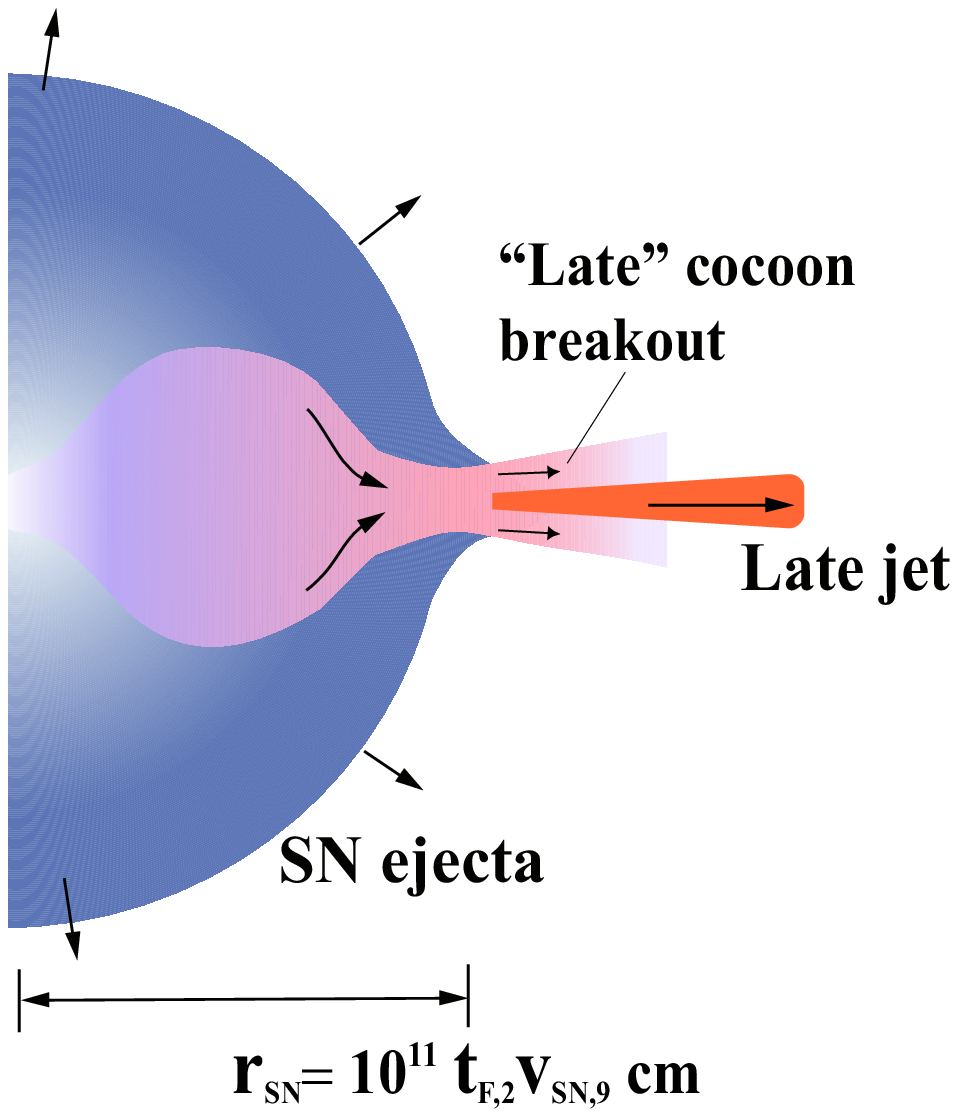}\\
\textbf{(a)}\\
\includegraphics[height=6cm, angle=0]{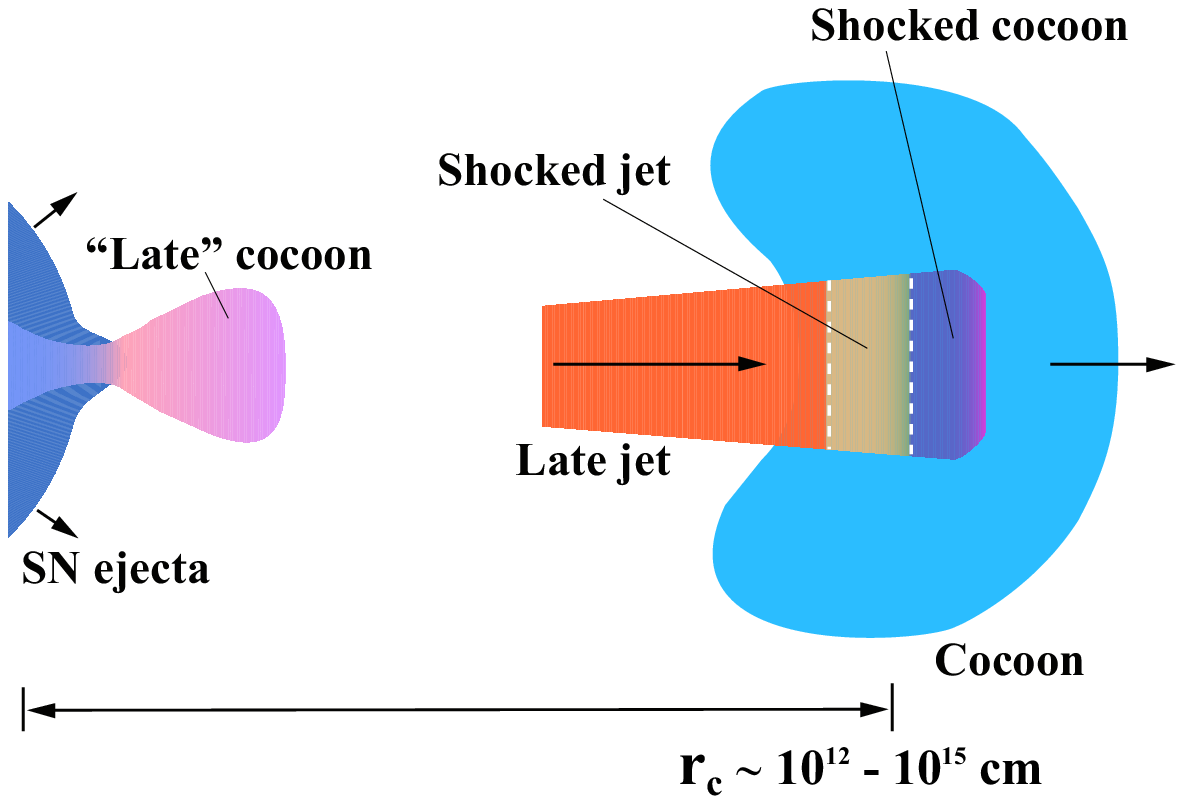}\\
\textbf{(b)}\\
\caption{\textbf{(a)}: Schematic illustration of the
late jet - SNa ejecta interaction, at a slightly later time than shown in Fig. \ref{fig:jet-illus}.
The breakout of the ``late'' cocoon produces a short thermal emission lasting for $\sim 10$ s.
\textbf{(b)}: Schematic illustration of the late jet - cocoon interaction, at a somewhat later
time than in (a), when the late jet has completely emerged out of the SNa ejecta and is colliding
with the cocoon created by the main GRB jet as it made its way through the progenitor star.
A pair of shocks are going through the cocoon and the late jet, respectively. Note that the
``late'' cocoon has expanded and cooled down at this time.}
\label{fig:interact-illus}
\end{figure}

The late jet will run into the SNa ejecta first (at a
distance $\sim 10^{11}$ cm; see Fig. \ref{fig:interact-illus}a),
and then catches up and run into the cocoon
-- if it successfully crosses the ejecta -- at a larger distance
($\sim 10^{12} - 10^{14}$ cm; see Fig. \ref{fig:interact-illus}b).


\section{Late jet - SNa ejecta interaction}\label{sec:jet-ejecta}

Since the late jet is highly relativistic and the SNa ejecta is
sub-relativistic, the late jet will catch up with the SNa ejecta in a time
roughly equal to the distance of the SNa ejecta from the explosion center
(where the jet emerged), $r_{SN}$, divided by light speed $c$.
If the late jet is launched with a delay of $t_F$, then
$r_{SN} \approx V_{SN} t_F \approx 10^{11} V_{SN,9}t_{F,2}$ cm
(hereafter we use the convention $X_n = X/10^n$
unless specifically notified). At this time the radial width of the ejecta
is about the same size of the initial stellar envelope, $\D_{SN} \sim r_* \sim
10^{11}$ cm. The ejecta has hardly moved from
the initial position of the progenitor stellar envelope for a jet delay
$t_F \sim 100$ s. The particle density in the SNa ejecta is
$n_{SN} \simeq 10^{24}\, V_{SN,9}^{-2}t_{F,2}^{-2}M_{SN,1} \D_{SN,11}^{-1}$
cm$^{-3}$,
where $M_{SN}$ is in units of the solar mass, and it is
extremely optically thick. If there is any emission from
the interaction between the late jet and the SN ejecta, that emission
should be thermal.

\subsection{The cavity in the SNa ejecta} \label{sec:cavity-ejecta}

We recall that the main GRB jet has already traversed the star and left
a cavity in the polar region before the late jet comes. Since the
material enclosing the ``wall'' of the cavity was heated up by the passage
of the main jet, it has tendency to refill the cavity. To find out whether the cavity
in the SNa ejecta has been filled up before the late-jet encounter, we
estimate the time it takes for filling up the cavity.

When inside the star, the cocoon material has a relativistic temperature,
i.e., the local sound speed $c_{s,c} = c/\sqrt{3}$.
Thus the cocoon material may fill the cavity on a time scale of
$r_* \theta_j/c_{s,c} \sim$ 1 s, much shorter than the onset  of
the late jet $t_F$. However the cocoon will also break out and flow
away from the ejecta at that same speed in $\sim$ 10 s,
leaving behind a somewhat evacuated polar region. The filling of the
polar cavity by the rest of the SN ejecta is uncertain.
Assuming a temperature $T \sim 10^8$ K for the SNa-shocked ejecta
material, it has $c_s \sim 1.2\times 10^8 T_8^{1/2}$
cm s$^{-1}$, so the filling time would be
$r_*\theta_j/c_s(T) \approx 10^2 r_{*,11}T_8^{-1/2}$ s, comparable to
$t_F$.
Considering that the ejecta local temperature possibly
decreases from inner parts to outer parts and the transverse size of
the cavity gets bigger outward, it is likely that when the late
jet hits the ejecta the cavity is partly filled -- the inner part is
filled but the outer part is not.
However, for a very late jet, e.g.,
$t_F \sim 10^3$ s, the cavity is surely filled before the late jet comes up; and then the
question that arises is whether the jet is powerful enough to cross
the new refilled cavity. This issue will be discussed elsewhere.

\subsection{Late jet - SNa ejecta crossing}

The late jet comoving particle density is $n_j = L_j/(\pi \G_j^2 c^3 r_{SN}^2
\theta_j^2 m_p) = 7\times10^{16} L_{j,49}\G_{j,2}^{-2}r_{SN,11}^{-2}\theta_{j,-1}^{-2}$ cm$^{-3}$,
where $L_j$ is the late jet luminosity. Because $n_{SN}/n_j \gg 1$, the
late jet will be decelerated to a non-relativistic speed after it
first hits the ejecta. It will undergo the same process as the GRB
main jet did when propagating through the progenitor star. A cocoon
forms in the SNa ejecta and makes a way for the jet head by pushing
the ejecta material sideways. To distinguish it from the cocoon
associated with the main GRB jet, let us call this cocoon associated
with the late jet as the ``late cocoon'' (see in Figs. \ref{fig:jet-illus}
and \ref{fig:interact-illus}a).

When the jet is moving inside the ejecta, the jet head has been slowed down to
be at a sub-relativistic speed (Ramirez-Ruiz et al. 2002; Matzner 2003):
\begin{displaymath}
v_h = \left( \frac{L_j}{\pi r_{SN}^2 \theta_j^2
\rho_{SN}c} \right)^{1/2}
\end{displaymath}
\begin{equation}\label{eq:v_h}
\;\;\;\;\;= 0.8\times 10^9
L_{j,49}^{1/2} \theta_{j,-1}^{-1} M_{SN,1}^{-1/2} \D_{SN,11}^{1/2}
\,{\rm cm\,s}^{-1}.
\end{equation}
 Note that if the SNa ejecta width is constant then $v_h$ has no dependence
on $r_{SN}$ and therefore on $t_F$; but if the SNa ejecta is uniformly
distributed from the centre to the radius $r_{SN}$, i.e., $\D_{SN} \approx
r_{SN}$, then $v_h$ increases with $t_F$ as $\propto t_F^{1/2}$.
SNa explosion simulations show the latter case, i.e., $\D_{SN} \approx
r_{SN}$, is the most probable one (e.g., Tanaka et al. 2009).

A constraint on the late jet property can be derived from the
requirement that the duration $t_{dur}$ of the jet must be larger
than the time that the jet spends to cross the SNa ejecta.
Thus $t_{dur} > \D_{SN}/v_h$ implies $L_{j,49}^{1/2}t_{dur,2}\theta_{j,-1}^{-1}
> 1.2\,M_{SN,1}^{1/2} \D_{SN,11}^{1/2}$.

\subsection{Thermal emission from the late cocoon break out} \label{sec:the-emi}

The late cocoon -- formed by the interaction
of late jet with SNa ejecta\footnote{The late cocoon should not be confused with the cocoon that
was formed and left behind by the main GRB jet. The latter will be discussed
in Section \ref{sec:jet-cocoon}.} -- will break out of the star immediately following
the breakout of head of the late jet (Fig. \ref{fig:interact-illus}a).
The luminosity of the thermal emission from the late cocoon breakout can be estimated
from its temperature and transverse size. Prior to its breakout, the late cocoon has a pressure
$p_c$ and its leading head moves with the jet head at the same speed in the radial
direction and expands transversely into the SNa ejecta with a speed $v_{\bot}$.
The ram pressure balance at the lateral interface between the late cocoon
and the ejecta material gives
\begin{equation} p_c = \rho_{SN} v_{\bot}^2,
\label{eq:p-lat}
\end{equation}
where $\rho_{SN}$ is the ejecta mass density.

The late cocoon contains an energy $E_c$ that is approximately equal
to the jet luminosity $L_j$ times the shell crossing time $\D_{SN}/v_h$, and
it is radiation pressure dominated. The volume of the late cocoon just before
the breakout is
$V_c= \pi \D_{SN}r_{\bot}^2 / 3$, where $r_{\bot}= \D_{SN} v_{\bot}/v_h$ is
the transverse size. So the pressure is
\begin{equation}\label{eq:p_c}
p_c = \frac{E_c}{3V_c} = \frac{L_j v_h}{\pi
\D_{SN}^2 v_{\bot}^2}.
\end{equation}
Combining with Eq. (\ref{eq:p-lat}) gives
\begin{equation}
p_c = \left(\frac{L_j \rho_{SN} v_h}{\pi \D_{SN}^2}\right)^{1/2},
\end{equation}
and the thermal temperature is
\begin{displaymath}
T_{th} = \left(\frac{3 p_c}{a}\right)^{1/4}
\end{displaymath}
\begin{equation}
~~~~~~ = 1.2\times10^8 \,
L_{j,49}^{3/16} \theta_{j,-1}^{-1/8} M_{SN,1}^{1/16} r_{SN,11}^{-1/4}
\D_{SN,11}^{-5/16} \,\, {\rm K},
\end{equation}
where Eq. (\ref{eq:v_h}) is used.
For a late jet with $t_F \sim 10^2$ s, the typical thermal
photon energy should be a few keV. $T_{th}$ becomes smaller for larger
$t_F$; for instance, when the SN ejecta width $\D_{SN}$ is $\approx r_{SN}$ and, if
$L_j \propto t_F^{-1.5}$ (Lazzati et al. 2008), we have $T_{th} \propto t_F^{-0.8}$.

Let us estimate the luminosity of the thermal emission from the late cocoon breakout.
The late cocoon transverse expansion speed is
\begin{displaymath}
v_{\bot}=\left(\frac{p_c}{\rho_{SN}}\right)^{1/2}
\end{displaymath}
\begin{equation}
~~~~~ = 0.6\times10^9  \, L_{j,49}^{3/8} \theta_{j,-1}^{-1/4} M_{SN,1}^{-3/8} r_{SN,11}^{1/2} \D_{SN,11}^{-1/8}\,
{\rm cm\,s}^{-1}.
\end{equation}
The cocoon transverse size is
$r_{\bot}= \D_{SN} v_{\bot}/v_h= 0.8\times10^{11} \, L_{j,49}^{-1/8} \theta_{j,-1}^{3/4} M_{SN,1}^{1/8} r_{SN,11}^{1/2} \D_{SN,11}^{3/8}\, {\rm cm}.$
Notice that $r_{\bot}$ is almost $\sim r_{SN}$ for fiducial parameter values.
However, the transverse size of the visible emitting region of
the cocoon just at the breakout should be smaller than
$r_{\bot}$. This is because when the jet head moves near
the outer surface of the ejecta, it probably accelerates due to the rapid
density drop of the stellar materia there; thus the lateral expansion of the
cocoon's leading head immediately following the jet head should be suppressed. Therefore
at that time, the overall cocoon might be in an ``hourglass''
shape, as illustrated in Ramirez-Ruiz et al. (2002) and in our
Fig. \ref{fig:interact-illus}a, rather than a conic shape. At the breakout, the hot cocoon
material escapes and accelerates radially from a ``nozzle'' which has a transverse
size on the same order of the jet's.
Thus, the transverse size of the visible emitting cocoon right at the
breakout can be estimated as
$r_{th, \bot}= r_{SN}\th_j= 10^{10}\, \th_{j,-1}V_{SN,9} t_{F,2}$ cm, while $r_{\bot}$
should be the transverse size of cocoon at its \textit{widest} cross section.

The luminosity of the black body radiation is
\begin{displaymath}
L_{th}= \sigma T_{th}^4 \pi r_{th, \bot}^2
\end{displaymath}
\begin{equation} \label{eq:lum-bb}
~~~~~~~ = 4.5\times10^{48} L_{j,49}^{3/4} \theta_{j,-1}^{3/2} M_{SN,1}^{1/4} r_{SN,11} \D_{SN,11}^{-5/4}\,
{\rm erg\,s}^{-1}.
\end{equation}
Using the fact that $L_j \propto t_F^{-1.5}$, one can find this thermal luminosity decreases with $t_F$ as $\propto t_F^{-1.4}$
if $\D_{SN} \approx r_{SN}$, or as $\propto t_F^{-0.13}$ if $\D_{SN} \sim$ constant. Thus, the radiation efficiency increases as $\propto t_F^{0.1}$ for $\D_{SN} \approx r_{SN}$ and as $\propto t_F^{1.4}$ for $\D_{SN} \sim$ constant, respectively, which is simply because the emitting transverse area increases with $t_F$.

This thermal transient will last for a time comparable to the
time it takes for the bulk of the late cocoon to escape the
ejecta. After that, the luminosity drops quickly due to adiabatic cooling.
The late cocoon's outflow speed is $\sim c_s= c/\sqrt{3}$,
so the escape time is $\D t_{esc} \approx \D_{SN}/c_s= 6 \, \D_{SN,11}$ s.
Note that $L_{th}$ is large because all the jet power was deposited in the
cocoon over a relatively long time (ejecta crossing time $\sim 10^2$ s), and
this energy is efficiently radiated away (via black body radiation) within a relatively
short time, $\sim 6$ s.

An even stronger ($L \sim 10^{49}$ erg s$^{-1}$)
thermal pulse is  associated with the main GRB jet break out. However, this thermal transient is short ($\sim$ 10 s), with spectral peak at X-rays, and it happens during the initial stage of the main ``burst'' $\g$-ray emission when the X-Ray Telescope (XRT) is not pointing towards the burst. Since this emission can be estimated to be one to two orders of magnitude dimmer than the emission of the burst itself, it is difficult to observe. On the other hand, the thermal transient due to the late cocoon breakout that we consider here arises later and at the time that the XRT is already pointing towards the burst. Hence this transient should be easier to detect.

The fact that we don't observe such thermal transient, preceding flares during the afterglow is somewhat puzzling.
If  the late jet is fairly early (say, $t_F \le 10^3$ s), the polar cavity in the SNa
ejecta left by the main GRB jet was probably only partially filled, i.e., only the
inner part of the cavity is filled (see Sec. \ref{sec:cavity-ejecta}). In that case,
the luminosity of the thermal emission would be much smaller than what we estimated
above. The situation is more problematic for $t_F \ge 10^3$.
We discuss further the observational prospects of this transient in Sec. \ref{sec:obs-jet-sn}.


\section{Late jet - cocoon interaction} \label{sec:jet-cocoon}

We now turn to the interaction of the late jet,
after it has successfully crossed the SNa ejecta, with the cocoon that was formed by
the main GRB jet (see Fig. \ref{fig:interact-illus}b).
The cocoon breaks out from the star at the same time when the main
jet breaks out. Then it accelerates to a mildly relativistic speed.
The delay of the late jet with respect to the cocoon breakout
is $t_F$ and the late jet catches up with
the adiabatically cooled cocoon at
$r_i \simeq c t_F /(\beta_j-\beta_c)$, where $\beta_j$ ($\G_j$) and $\beta_c$
($\G_c$) are the speed (LFs) of the late jet and the cocoon,
respectively. When the cocoon speed is mildly relativistic,  $r_i
\approx 6 \times 10^{14} t_{F,2}\G_{c,1}^2$ cm, which is much further than the late
jet - SNa ejecta interaction site. For a sub-relativistic cocoon,
$r_i$ is close to but still outside the  jet - SNa ejecta interaction region.

\subsection{Cocoon geometry and dynamics}

At the breakout, the cocoon has an energy $E_c$, and an energy-to-mass
ratio $\et_c$. The cocoon opening angle $\th_c$ is
determined by its transverse expansion speed $\sim c/\sqrt{3}$, thus
$\theta_c \sim 1/\sqrt{3} = 0.6$. During the early stages of expansion,
the cocoon's radial width, $\Delta_c$, is approximately the width of the stellar
envelope $r_*$. Later on, due to the radial expansion of a relativistically
moving gas, $\D_c$ asymptotically approaches $r/(2\G_c^2)$ in the lab frame; this
happens when $r \ge r_w \approx r_* \et_c^2$. The cocoon's LF can be
described as $\G_c (r) \approx \th_c r/ r_*$ when $r < r_s$, and $\G_c \approx \et_c$
when $r \ge r_s$, where $r_s= \et_c r_*/\th_c$ is the saturation radius
(Paczy\'{n}ski 1986; Goodman 1986; Shemi \& Piran 1990; Piran, Shemi \& Narayan 1993;
M\'{e}sz\'{a}ros, Laguna \& Rees 1993). The evolution of the cocoon's comoving volume,
$V_c'(r) = \pi \th_c^2 r^2 \D_c \G_c(r)$, is described by
\begin{equation}
V_c'=\left\{ \begin{array}{ll}
\pi \th_c^3 r^3, & \,\,\, {\rm for} \,\,\, r < r_s\\
\pi \th_c^2 r^2 \et_c r_*, & \,\,\, {\rm for} \,\,\, r_s < r < r_w\\
\pi \th_c^2 r^3 /(2\et_c), & \,\,\, {\rm for} \,\,\, r > r_w.
\end{array} \right.
\end{equation}

The pressure evolution of the cocoon follows the adiabatic expansion law:
$p_c \propto V_c^{-\g}$. Initially the radiation pressure dominates, so $\g=4/3$.
When the cocoon's optical depth to Thomson scattering decreases to below unity,
the photons decouple from the matter and the radiation pressure drops exponentially;
then the gas pressure takes over the dominance of the pressure with $\g= 5/3$. The transition happens at
radius $r_t = [\sigma_T E_c / (\et_c m_p c^2 \pi \th_c^2)]^{1/2}
= 3.5\times10^{14} E_{c,51}^{1/2} \et_{c,1}^{-1/2}$ cm, where $\sigma_T$ is the
Thomson scattering cross section.
The initial pressure at the breakout is given by
\begin{equation}\label{eq:ini-pre-coc}
p_{c,0}= \frac{E_c}{3 V_c' (r_*)} = 5\times10^{17} E_{c,51} r_{*,11}^{-3} \,\,{\rm dyn}\,\,{\rm cm}^{-2} .
\end{equation}
The evolution of the cocoon pressure is thus given by
\begin{equation}\label{eq:pre-evo-coc}
\frac{p_c(r)}{p_{c,0}}= \left\{ \begin{array}{ll}
\left(\frac{r_*}{r}\right)^4, & \,\,\, {\rm for} \,\,\, r < r_s,\\
\left(\frac{\th_c}{\et_c}\right)^{4/3} \left(\frac{r_*}{r}\right)^{8/3}, & \,\,\, {\rm for} \,\,\, r_s < r < r_w,\\
(2\et_c\th_c)^{4/3} \left(\frac{r_*}{r}\right)^4, & \,\,\, {\rm for} \,\,\, r > r_w.
\end{array} \right.
\end{equation}

The comoving density of the cocoon is assumed to be homogeneous; the same is for
the late jet. The width of the late jet is determined by its duration and the
radial expansion, so $\Delta_j(r)= c t_{dur} +  r/(2\G_j^2)$.
The jet comoving density is
\begin{equation}
n_j(r)= \frac{E_{j,iso}}{4\pi r^2 m_p c^3 \G_j^2 (t_{dur}+\frac{r}{2\G_j^2c})}.
\end{equation}

\subsection{Cavity in the cocoon} \label{sec:cavity-cocoon}

There was also initially a cavity in the cocoon left by the passage of the main jet.
Here we estimate how quickly the cavity would be filled.
The filling up process starts when the main jet dies off. For a typical duration
of GRB $t_{grb} \approx 10$ s, the cocoon has moved to a radius
$r_c \approx c t_{grb} = 3\times10^{11} \,\, t_{grb,1}$ cm. Since $r_c < r_s$,
the cocoon gas is still relativistic, the sound speed is $c_s \approx c/\sqrt{3}$,
and the time required for cavity to close is
$\approx r_c \th_j/c_s = 0.6 \,\, t_{grb,1} \th_{j,-1}$ s
which is $\ll t_F$. Thus the cavity is securely filled when the late jet reaches the cocoon.

Now let us consider a possibility that there is a continuous low-level central
engine activity (a {\it low-power} jet) following the end of the main GRB jet and preceding
the late jet. This low-power jet and the late jet more or less are the same phenomenon,
only with different energy fluxes. The low-power jet might have too small radiation
luminosity to have an observational imprint, however it might still be dynamically
important for keeping open the polar cavity in the SNa ejecta and the cocoon.

To keep the cavity open without significant energy dissipation from the jet and cocoon,
the transverse pressure, i.e., the thermal pressure, of the low-power jet should be
greater than or equal to the cocoon's pressure at $r_c$.
Though the low-power jet started with a high thermal pressure at a
distance $r_0 \sim 10^2$ km from the explosion centre where it was launched, at $r_c$ its thermal
pressure has dropped to be much smaller than the cocoon pressure because it has adiabatically
expanded by a much larger factor than the cocoon does. Thus, when this low-power jet entered
the cavity left by the main GRB jet, it will be squashed rapidly by
the gas pressure in the cocoon.

The ram pressure of the low power jet, $p_{j,ram} (r_c) = L_{j,low}/(4\pi r_c^2 c)$ where $L_{j,low}$
is the jet luminosity, can help bore a
cavity through the cocoon under suitable conditions. For a long-lasting, continuous jet, the event that takes place
after it has been squashed is as follows. If $p_{j, ram}(r_c) > p_c(r_c)$, then the residual,
incoming jet will punch a new channel through the cocoon. In doing this, the jet is heated up by the
reverse shock, so that the jet which is moving inside the channel could have an enhanced
$p_j$ (thermal pressure) that is comparable to $p_c$, and can keep the channel open. If we assume
the low-power jet was launched at the end of the GRB, i.e., $t_{grb} \sim 10$ s, making the equality
$p_{j, ram}= p_c(r_c)$ gives $L_{j,low}= 2\times10^{48}$ erg s$^{-1}$, with other parameters at
fiducial values. The same equality at later times gives the time dependence
$L_{j,low}(t) \propto t^{-2}$. So a decaying luminosity profile $L_{j,low}(t)= 2\times10^{48} t_1^{-2}$
erg s$^{-1}$ for the low-power jet is required to keep the cavity open, i.e., the minimum required
total energy for a low-power jet to keep the cavity open is a few $\times 10^{49}$ erg.
Note that at this minimal jet luminosity, the process of keeping the cavity open
is {\it not} smooth and a fraction of jet energy is dissipated and it should have some radiation associated with it.

In conclusion, the polar cavity left by the main GRB jet in the cocoon can fill up quickly -- before the
late jet reaches the cocoon. The presence of a continuous low-power jet can keep this cavity open, provided that
the jet has a minimal total energy of a few $\times 10^{49}$ erg. In the following calculation
we consider the case when the cavity is filled up. Clearly if the cavity is empty the late jet cocoon
interaction is trivial.

\begin{figure*}
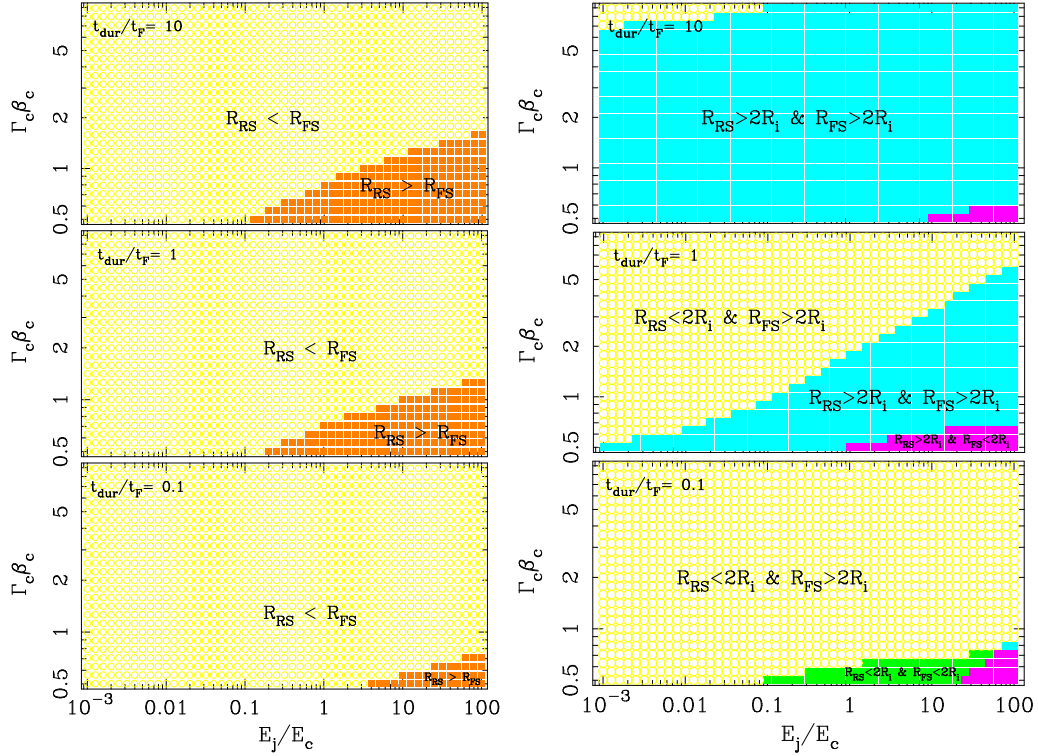

\centerline{
\includegraphics[width=10cm,angle=270]{sho_cro_t_dep.ps}
\hspace{0.1cm}
\includegraphics[width=10cm, angle=270]{sho_cro_t_dep_rrs_rfs_ri.ps}}
\caption{{\it Left} panels: the comparison between the RS crossing radius $r_{RS}$
and the FS crossing radius $r_{FS}$ in the late jet - cocoon interaction for
a set of model parameter space. {\it Right} panels: the comparison of the two crossing
radii with $2 r_i$, where $r_i$ is the radius at which
the late jet catches up with the cocoon and the interaction begins. $\G_c$
and $\b_c$ are the cocoon's LF and dimensionless speed, respectively. $E_j$ and
$E_c$ are the kinetic energies of the late jet and the cocoon, respectively. Both
crossing radii are numerically calculated from Eq. (\ref{eq:sho-cro}). Other parameter values
are: $E_c= 10^{50}$ erg, $\th_j=0.1$, $\th_c=0.6$, $t_F= 300$ s, $\G_j= 100$,
$\eps_e=0.1$ and $\eps_B=0.01$. We find the results in this
figure do not depend on $t_F$ or $\G_j$ as long as $t_F \ge 10^2$ s, and
$\G_j \gg 1$ (say $\sim 10^2$).}
\label{fig:r_RS-r_FS}
\end{figure*}


\subsection{The jet - cocoon  interaction phases}

The dynamical process of the late jet - cocoon interaction can be
decomposed into three phases in the following sequence\footnote
{The three-phase decomposition treatment follows Dermer (2008)
who studied the emission due to the external shocks
between a GRB jet and a stationary circumburst dense cloud.}: (i) The
{\it Collision Phase} takes place when the jet runs into
the cocoon with a forward shock (FS) propagating into the cocoon and
a reverse shock (RS) propagating into the jet (see Fig. \ref{fig:interact-illus}b).
(ii) The {\it Penetration Phase} begins when either the
RS crosses the entire jet or the FS crosses the entire cocoon,
whichever comes first. In the first case
the shocked fluid (RS-shocked jet and FS-shocked cocoon fluid) will
decelerate, , after RS crosses the jet, when more and more cocoon material are swept by the FS.
In the latter case the shocked
fluid will be accelerated, after FS crosses the entire cocoon, by the remaining unshocked jet ejecta, and a new
particle population will be accelerated at the RS. (iii) The {\it
Expansion Phase} begins when both the FS and RS have run through
their courses, and the entire shocked fluid expands adiabatically.

\subsubsection{The collision phase}\label{col_pha}

In this phase, the FS propagates into the cocoon and the RS
propagates into the jet. For simplicity we approximate the interaction using
a planar geometry. The entire jet / cocoon system can be
divided into several zones. Outside the FS (RS) front is the unshocked
cocoon (jet), and these are taken to be cold plasma, i.e., $e= n$ and
$p= 0$, where $e$, $p$ and $n$ are the fluid energy density
(including the rest mass energy), pressure and
particle number density, respectively, all measured in its comoving frame.
In between the FS front and the RS front are the shocked cocoon fluid and
the shocked jet fluid. These shocked fluids move with the same LF ($\G_{s}$)
and have the same thermal pressure; they are separated by a contact
discontinuity (CD) plane.

The fluid properties across a shock front are governed by the mass, momentum
and energy conservation laws (e.g., Landau \& Lifshitz 1959;
Blandford \& McKee 1976). Across the shock the fluid particle density increases
by a factor of $(\hat{\g}\bar{\G}+1)/(\hat{\g}-1)$,
and $e=\bar{\G} n$, where $\bar{\G}$ is the shocked fluid
LF measured in the unshocked fluid comoving frame; $\hat{\g}$ is given by:
$p=(\hat{\g}-1)(e-n)$. We use $m_p=c=1$ to simplify the formulae.

At the CD, the pressure in the shocked jet material equals the pressure in
the shocked cocoon material, i.e.,
\begin{displaymath}
(\hat{\g}_{RS}-1) (\bar{\G}_{sj}-1) \left(\frac{\hat{\g}_{RS} \bar{\G}_{sj}+1}{\hat{\g}_{RS}-1}\right) n_j
\end{displaymath}
\begin{equation} \label{eqn:pre-bal}
~~~~~~~~~~~~~~~~~ = (\hat{\g}_{FS}-1) (\bar{\G}_{sc}-1) \left(\frac{\hat{\g}_{FS} \bar{\G}_{sc}+1}{\hat{\g}_{FS}-1}\right) n_c,
\end{equation}
where the subscript ``RS'' refers to the reverse-shocked fluid and ``FS'' the
forward-shocked one, and $\bar{\G}_{sj}$ and $\bar{\G}_{sc}$ are the shocked fluid
LFs measured in the unshocked jet and cocoon comoving frames, respectively.
$\hat{\g}$ lies between 4/3 and 5/3, and can be written in terms of $\bar{\G}$
as $\hat{\g}= (4\bar{\G}+1)/3\bar{\G}$ (Kumar \& Granot 2003).
Then Eq. (\ref{eqn:pre-bal}) simplifies to
\begin{equation}
(\bar{\G}_{sj}^2-1)n_j= (\bar{\G}_{sc}^2-1)n_c.
\end{equation}
Since $\bar{\G}_{sj}=\G_j\G_s(1-\b_j\b_s)$ and $\bar{\G}_{sc}=\G_s\G_c(1-\b_s\b_c)$,
we find from the last equation that
\begin{equation}
\G_s= \G_j \frac{\sqrt{a}+\G_c/\G_j}{(a+1+2\sqrt{a} \bar{\G}_{jc})^{1/2}},
\end{equation}
where $a= n_j/n_c$ is the density ratio, $\bar{\G}_{jc}$ is the unshocked jet LF
measured in the unshocked cocoon rest frame.
Note that this expression for $\G_s$ is valid for both sub-relativistic and
relativistic shocks, and for $4/3 \le \hat{\g} \le 5/3$.

The shock (RS or FS) front moves with a LF different from the LF of the
shocked fluid. The shock front LF as measured in the unshocked fluid comoving frame --
which we denote as $\bar{\G}_{RS, j}$ for the RS and as $\bar{\G}_{FS, c}$ for the FS --
is given by the solution to the shock-jump conditions as a function of $\hat{\g}$
and $\bar{\G}$ (Eq. 5 of Blandford \& McKee 1976). Using the expression of $\hat{\g}$
in terms of $\bar{\G}$, we find
\begin{equation}
\bar{\G}_{RS, j} = \frac{4\bar{\G}_{sj}^2-1}{\sqrt{8\bar{\G}_{sj}^2+1}},
\;\;\; {\rm and} \;\;\;
\bar{\G}_{FS, c} = \frac{4\bar{\G}_{sc}^2-1}{\sqrt{8\bar{\G}_{sc}^2+1}}.
\end{equation}
These expressions are valid for both sub-relativistic and
relativistic shocks.

The pair of shocks exist until one of the two shocks, either the RS or FS, has traversed
through the unshocked fluid. From that time on, the interaction
will move to the next dynamic phase ({\it Penetration}). To determine which shock
(RS or FS) crossing
occurs first, we calculate two radii, $r_{RS}$ and $r_{FS}$, where $r_{RS}$
is the distance of the system when the RS crosses the rear end of the jet, and $r_{FS}$
is when the FS crosses the front end of the cocoon, pretending that the pair of shocks
had existed all the way to the larger of the two radii.

At $r_{RS}$ or $r_{FS}$, the total distance that the shock has traveled through the
unshocked fluid is equal to the radial width of the jet or the cocoon at that radius.
Thus the two radii can be obtained by solving the equations
\begin{displaymath}
\D_j(r_{RS})= \int_{r_i}^{r_{RS}} \frac{(\b_j-\b_{RS})}{\b_j} d r \;\;\;\;
{\rm and} \;\;\;\;
\end{displaymath}
\begin{equation}\label{eq:sho-cro}
~~~~~~~~~~~~~~~~~~~~ \D_c(r_{FS})= \int_{r_i}^{r_{FS}} \frac{(\b_{FS}-\b_c)}{\b_c} d r ,
\end{equation}
where we have used $dt= dr/\b_j \simeq dr/\b_c$.

If $r_{RS} < r_{FS}$,  the RS crosses the jet before the FS crosses the cocoon,
and vice versa.  We calculate $r_{RS}$ and $r_{FS}$ for different parameters.
The results are shown in Fig. \ref{fig:r_RS-r_FS}. We find that
$r_{RS} < r_{FS}$, i.e., the RS crossing occurs first, for most of the parameter space;
$r_{RS} > r_{FS}$ can only happen when the cocoon bulk motion is sub-relativistic ($\G_c\b_c < 1$)
and the energy carried by the late jet is much larger than that of cocoon ($E_j/E_c \gg 1$).
We will use this result later (Sub-sections
\ref{sec:emi-jet-coc} - \ref{sec:lc-shape}) to simplify the calculation of the light curve by assuming that RS crossing always occurs before
FS crossing.


\subsubsection{The penetration phase}

After the RS crosses the jet, the FS continues to pass through the cocoon.
We consider the entire shocked fluid, both the old and the newly shocked,
moving together with the same LF. In the rest frame of the
unshocked cocoon, the LF of the shocked fluid is determined by the equation for
the deceleration of a relativistic blast wave in the adiabatic limit as the
blast wave sweeps up the stationary ambient medium:
\begin{equation}\label{B&D00}
\G_s'(x')= \frac{\G_{s,\D}'}{\sqrt{1+2\G_{s,\D}'^2 m_c(x')c^2/E_{j,iso}'}}
\end{equation}
(B\H{o}ttcher \& Dermer 2000) where the prime sign denotes the unshocked
cocoon rest frame, $\G_{s,\D}'$ is the shocked fluid LF at the end of the collision
phase, $m_c(x')$ is the isotropic equivalent swept-up mass at the distance $x'$
that the blast wave has traveled.

A significant deceleration of FS occurs after a point where
$\G_{s,\D}'^2 m_c(x_d') c^2 = E_{j,iso}'$. Before this point, the FS is coasting
into the unshocked cocoon at roughly the same speed it had prior to the RS
crossing. After this point, the FS decelerates as it sweeps more and more
cocoon material (similar to the external shock scenario for the GRB afterglows).
If the FS crossing is earlier than the RS crossing, the process is similar except
that it is the RS that continues to travel through the unshocked jet.

\subsubsection{The expansion phase}

After the FS eventually crosses the entire cocoon, the shocked fluid expands
outward with a LF determined by eq. (\ref{B&D00}) but with $m(x')$ replaced by the
isotropic equivalent total mass of the cocoon. The relativistic electrons
cool via radiation and expansion. The radiation from an adiabatically expanding
relativistic shell, when it is optically thin to the Thomson scattering, is discussed
in Barniol Duran \& Kumar (2009). We will address the optical thick case in
Sec. \ref{sec:opt-thick}.


\begin{figure}
\centerline{
\includegraphics[width=6cm,angle=270]{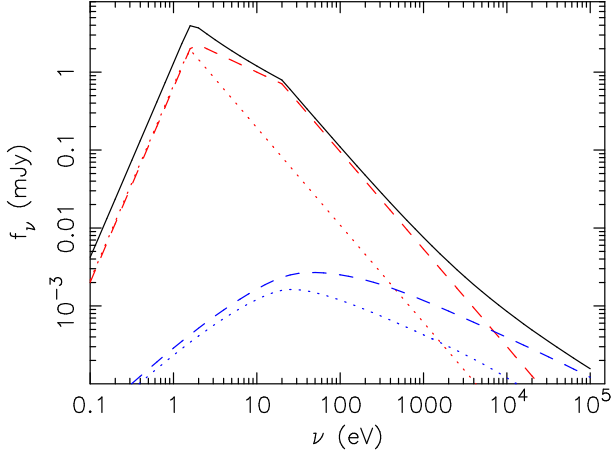}}
\caption{The observed spectrum from the late jet - cocoon
interaction at the peak of the light curve for $E_j=E_c=10^{50}$ erg,
$\G_c=3$, $t_{dur}/t_F= 0.3$ and $z= 2$. Other parameter values are same as
in Fig. \ref{fig:r_RS-r_FS}. Various lines represent contributions from
different emission regions and spectral components: {\it red dashed} - RS
synchrotron; {\it red dotted} - RS SSC; {\it blue dashed} - FS synchrotron;
{\it blue dotted} - FS SSC; {\it solid} - the sum.
For these parameter values, the order of the frequencies is
$\nu_c < \nu_{opt} < \nu_a < \nu_i < \nu_X$; $\nu_{opt}$ is in the
synchrotron-self-absorption optically thick regime, and the spectral
peak is around UV band. The SSC emission contribution is
important only for $\g$-ray band and above but is negligible at both
optical and X-ray bands.}
\label{fig:spec}
\end{figure}

\subsection{The emission from the late jet - cocoon interaction}\label{sec:emi-jet-coc}

We calculate the emission from the late jet - cocoon interaction,
and estimate the flux densities at the optical and X-ray bands using
standard shock synchrotron emission (e.g., Sari et al. 1998), and taking
into account the synchrotron self Compton (SSC) radiation.
Just behind the shock front, a fraction of the
bulk kinetic energy of the fluid upstream, $\eps_e$, is transferred to
the electrons, and another fraction, $\eps_B$, goes to the magnetic field.
All the swept-up electrons are shock-heated into a power-law energy
distribution with a spectral index $p$. The minimum LF of shock heated electrons is
\begin{equation}
\g_i = \eps_e \frac{m_p}{m_e}\left(\frac{p-2}{p-1}\right)(\bar{\G}-1),
\end{equation}
where $\bar{\G}$ is the shocked fluid LF measured in the unshocked fluid frame.
The magnetic field energy density downstream of the shock front is given by
\begin{equation}
U_B'= \frac{B'^2}{8\pi}= 4\bar{\G}(\bar{\G}-1)\eps_B n_0 m_p c^2,
\end{equation}
where $B'$ is the comoving frame field strength and $n_0$ is the particle number
density of the unshocked cocoon or jet. The synchrotron characteristic frequency
that corresponds to $\g_i$ is
\begin{equation}
\nu_i= \frac{eB'\g_i^2 \G_s}{2\pi m_e c (1+z)},
\end{equation}
where $e$ is the electron charge.

The electron cooling LF, $\g_c$, is determined by radiative
cooling due to synchrotron and SSC radiations:
\begin{equation} \label{eq:gam_c}
\g_c m_e c^2 = \frac{4}{3}\sigma_T c \g_c^2 U_B' (1+Y) t',
\end{equation}
where $\sigma_T$ is the electron's Thomson cross section, $t'$ is the
elapsing time in the shocked fluid rest frame, and $Y$ is
the Compton parameter defined as the ratio of the SSC to synchrotron
luminosities; $\g_c$ is obtained by numerically solving Eq. (\ref{eq:gam_c})
(e.g., McMahon, Kumar \& Piran 2006).
Electrons with LF $> \g_c$ will cool to $\g_c$ in time $t'$; the cooling
effect is negligible for electrons with LF $< \g_c$. The synchrotron
cooling frequency is
\begin{equation}
\nu_c = \frac{\G_s eB'\g_c^2}{2\pi m_e c (1+z)}.
\end{equation}

The self-absorption frequency for the synchrotron electrons, $\nu_a'$,
measured in the shocked fluid comoving frame is calculated by
(Sari \& Piran 1999; Li \& Song 2004;, McMahon et al. 2006; Shen \& Zhang 2009)
\begin{equation}
\max(\g_m, \g_a)\times2m_e \nu_a'^2=  F_{\nu_a'}',
\end{equation}
where $F_{\nu_a'}'$ is the flux density at $\nu_a'$
radiated away from the surface of the shocked region.

The emergent synchrotron spectrum of the shock-heated electrons can be
approximated as a piece-wise power law function. The peak of the
$f_{\nu}$-spectrum is at $\nu_{max} = \min(\nu_i, \nu_c)$ and the flux
density at the peak is
\begin{equation}
f_{\nu, max}= \frac{N_e}{4\pi D^2}\frac{\G_s m_e c^2 \sigma_T B'}{3e(1+z)},
\end{equation}
where $N_e$ is the isotropic equivalent total number of shock-heated
electrons, $D$ is the luminosity distance. We also calculate the
observed flux density due to SSC by (Rybicki \& Lightman 1979)
\begin{displaymath}
f^{ic}(\nu)= \frac{3}{4}\sigma_T \del s
\end{displaymath}
\begin{equation}
~~~~~~~~~~~~~ \times \int \frac{d\nu_s}{\nu_s^2}
\nu f^{syn}(\nu_s) \int_{\g_i}^{\infty}\frac{d\g}{\g^2}
n_e(\g)F\left(\frac{\nu}{4\g^2\nu_s}\right),
\end{equation}
where $\del s$ is the line-of-sight width of the emitting source,
$\nu_s$ and $f^{syn}(\nu_s)$ are the synchrotron frequency and flux
density (in the observer frame), respectively, and $n_e(\g)$ is the
number of shocked electrons per unit volume per unit interval of $\g$;
$n_e(\g) \propto \g^{-p}$ for $\g > \g_i$ [the modification of
$n_e(\g)$ due to the radiative cooling is included in our calculations].
The function $F(x)=2x \ln{x}+ x+1-2x^2$ for $0<x<1$
and is 0 otherwise. Using the expression for optical depth $\tau_e$ due
to Thomson scattering, the SSC flux density can be written as
\begin{displaymath}
f^{ic}(\nu)= \frac{(3/4)\tau_e}{\int_{\g_1}^{\infty}n_e(\g) d\g}
\end{displaymath}
\begin{equation} \label{eq:flux-ic}
~~~~~~~~~~~~~ \times \int \frac{d\nu_s}{\nu_s^2}
\nu f^{syn}(\nu_s) \int_{\g_1}^{\infty}\frac{d\g}{\g^2} n_e(\g)
F\left(\frac{\nu}{4\g^2\nu_s}\right).
\end{equation}

Fig. \ref{fig:spec} depicts the spectrum observed at the time
when the RS crosses the jet. The order of the characteristic frequencies
in this example is $\nu_c < \nu_{opt} < \nu_a < \nu_i < \nu_X$, and
$\nu_a$ is in the UV band. The SSC contribution is negligible
at optical band and {\it it} becomes important for photon energy $\ge \sim 1$ keV.


\begin{figure}
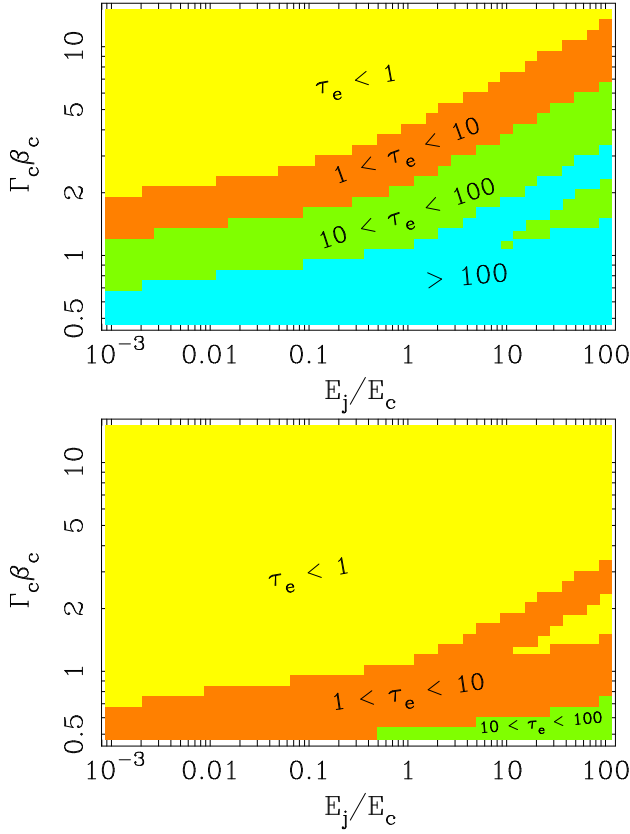

\centerline{
\includegraphics[width=5.5cm,angle=270]{tau_e_tf_100.ps}}
\centerline{
\includegraphics[width=5.5cm, angle=270]{tau_e_tf_1e3.ps}}
\caption{Contours of $\tau_e$ --- the shocked region optical depth to Thomson
scattering at the peak of the flux from the late jet - cocoon
interaction. {\it Top}: for $t_F= 10^2$ s; {\it Bottom}: for $t_F= 10^3$ s.
For even larger $t_F$ (say $\sim 10^4$ s), we find $\tau_e < 1$ for all
parameter space. Other parameter values are same as in Fig. \ref{fig:r_RS-r_FS}
except $t_{dur}/t_F= 0.3$.}
\label{fig:tau_e}
\end{figure}

\begin{figure}
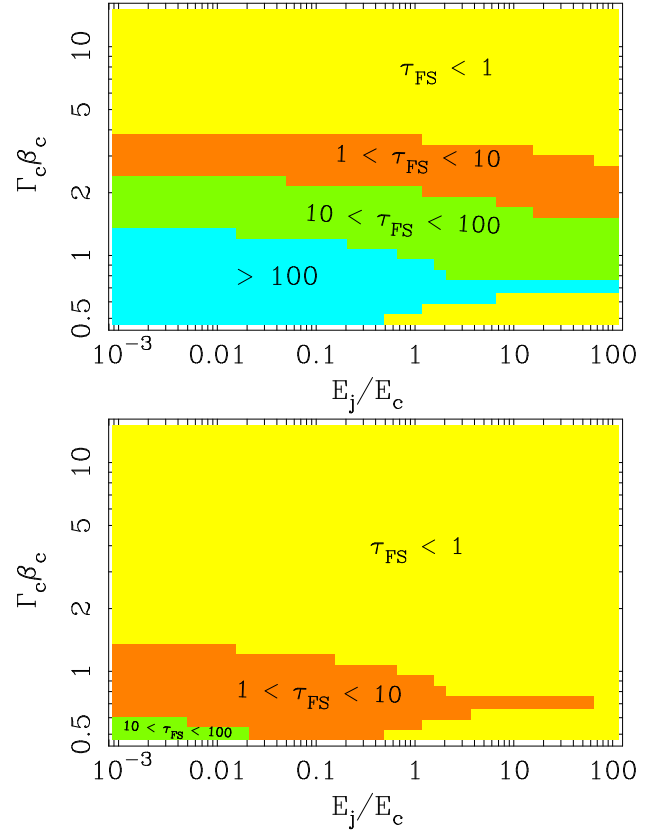

\centerline{
\includegraphics[width=5.5cm,angle=270]{tau_fs_tf_100.ps}
}
\centerline{
\includegraphics[width=5.5cm, angle=270]{tau_fs_tf_1e3.ps}
}
\caption{Contours of $\tau_{FS}$ --- the optical depth for the unshocked cocoon
that is still in front of the FS at the peak of the flux from the late jet -
cocoon interaction. {\it Top}: for $t_F= 10^2$ s; {\it Bottom}: for $t_F= 10^3$ s.
For even larger $t_F$ (say $\sim 10^4$ s), we find $\tau_{FS} < 1$ for all
parameter space. Other parameter values are same as in Fig. \ref{fig:r_RS-r_FS}
except $t_{dur}/t_F= 0.3$.}
\label{fig:tau_fs}
\end{figure}


\subsection{Light curves} \label{sec:lc-shape}

The light curve from the the late jet - cocoon interaction is
mainly determined by the evolution of $f_{\nu, max}$, $\nu_i$, $\nu_c$ and $\nu_a$.
The optical depth of the shocked fluid region may alter the final
light curve shape, which will be addressed later in this sub-section (\S \ref{sec:opt-thick}).
We follow the treatment of Yu \& Dai (2009) and calculate the light curve.
We define $T_{exp}$ as the
time when the shocked fluid has traveled a distance of $2 r_i$, where $r_i$
is the interaction radius (here and in the following, times denoted with the
capitalized letter ``T'' are the observer's times and $T= 0$ is the time when
the interaction begins). Thus, $T_{exp} = r_i/(2\G_s^2 c)$; before
$T_{exp}$, the increase of the radius can be neglected and $B'$
and $\G_s$ are considered to be constant; after $T_{exp}$, the attenuation of the
density and $B'$ due to the radius increase must be taken into account.

We also define the shock-crossing time $T_{cro}= \min(T_{RS}, T_{FS})$, where
$T_{RS}$ and $T_{FS}$, calculated in Eq. (\ref{eq:sho-cro}), are the crossing
times for the reverse shock and the forward shock, respectively
(Fig. \ref{fig:r_RS-r_FS} shows $T_{RS} < T_{FS}$ for most of the model
parameter space). Before $T_{cro}$, the radial spreading
of the shocked region is suppressed by the existence of two shocks, thus the
volume of the shocked region $V' \propto r^2$ and the internal energy density
$e' \propto V'^{-1} \propto r^{-2}$; the total number of shock heated particles
increases linearly with time. After $T_{cro}$, the radial expansion has to be
considered and the shocked region experiences adiabatic cooling. During this
phase, $V' \propto r^s$ (where $s= 2\sim3$) and the internal energy density
$e' \propto V'^{-4/3} \propto r^{-4s/3}$.

Therefore the evolution of $B'$ and $f_{\nu, max}$ are as follows:
(i) For $T_{\rm cro}<T_{\rm exp}$,
\begin{equation}
B'\propto\left\{\begin{array}{ll} T^0,~~~~~~~T<T_{\rm
exp};\\
T^{-2s/3},~~T>T_{\rm exp};
\end{array}\right.,
\end{equation}
\begin{eqnarray}
f_{\nu,\max}\propto\left\{\begin{array}{ll} T,~~~~~T<T_{\rm
cro};\\
T^0,~~~~~T_{\rm cro}<T<T_{\rm exp};\\
T^{-2s/3},~~T>T_{\rm exp};\end{array}\right.
\end{eqnarray}
(ii) For $T_{\rm cro}>T_{\rm exp}$,
\begin{equation}
B' \propto\left\{\begin{array}{ll} T^0,~~~~~~~T<T_{\rm
exp};\\T^{-1},~~~~~T_{\rm exp}<T<T_{\rm cro};\\
T^{-2s/3},~~T>T_{\rm cro};
\end{array}\right.
\end{equation}
\begin{eqnarray}
f_{\nu,\max}\propto\left\{\begin{array}{ll} T,~~~~~T<T_{\rm exp};\\
T^0,~~~~~T_{\rm exp}<T<T_{\rm cro};\\
T^{-2s/3},~~T>T_{\rm cro}.\end{array}\right. \label{fmax}
\end{eqnarray}

And the temporal dependences of the characteristic frequencies are given by
\begin{equation}
\nu_i \propto\left\{\begin{array}{ll} T^0,~~~~~T<T_{\rm exp};\\T^{-2s/3},~~T>T_{\rm
exp},\end{array}\right.
\end{equation}
\begin{equation}
\nu_c \propto\left\{\begin{array}{ll} T^{-2},~~~~T<T_{\rm exp};\\T^{2s-2},~~T>T_{\rm
exp},\end{array}\right.
\end{equation}
for $T_{\rm cro}<T_{\rm exp}$, and
\begin{equation}
\nu_i \propto\left\{\begin{array}{ll} T^0,~~~~~T<T_{\rm
exp};\\T^{-1},~~~~T_{\rm exp}<T<T_{\rm cro};\\
T^{-2s/3},~~T>T_{\rm cro},
\end{array}\right.
\end{equation}
\begin{equation}
\nu_c \propto\left\{\begin{array}{ll} T^{-2},~~~~T<T_{\rm
exp};\\T,~~~~~T_{\rm exp}<T<T_{\rm cro};\\
T^{2s-2},~~T>T_{\rm cro},
\end{array}\right.
\end{equation}
for $T_{\rm cro}>T_{\rm exp}$.

Note that the evolution of $\nu_c$ given above ignores SSC cooling.
SSC cooling is included in our numerical calculation as described
by Eq. (\ref{eq:flux-ic}). We find that the resultant $\nu_c$
scalings are not very different from those given above.

To calculate the light curve, we first calculate $B'$, $f_{\nu, max}$, $\nu_i$,
$\nu_c$ and $\nu_a$ at the expected peak time $T_p= \min(T_{cro}, T_{exp})$,
then we use the temporal evolution of $B'$ and $f_{\nu, max}$ to get
the observed flux density at other times. The SSC contribution to the flux
is included in the light curve calculation.
In addition, the optical thickness of the
late jet - cocoon system to Thomson scattering could alter the light curve shape because
it could delay the escape of the photons from the system; we will consider
this next, and then discuss the light curve results.


\begin{figure}
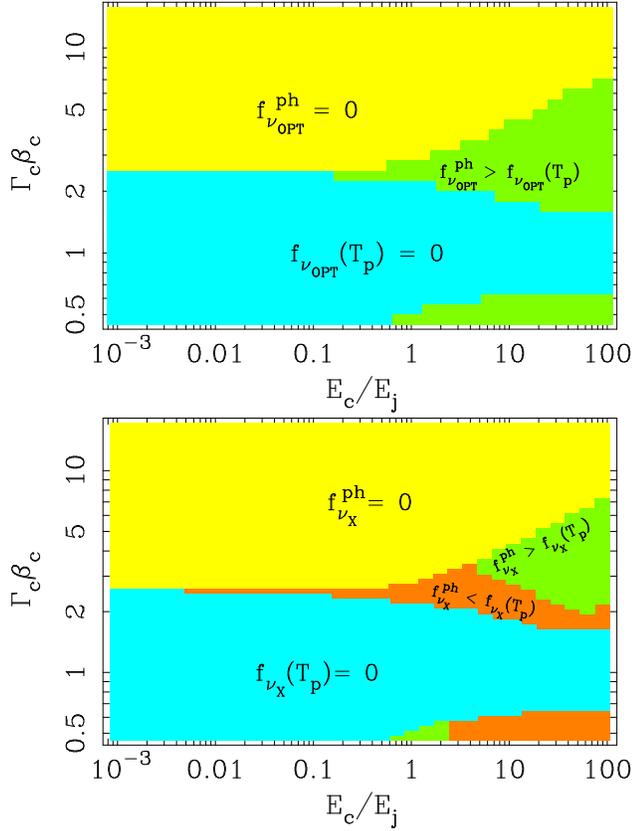

\centerline{
\includegraphics[width=5.5cm,angle=270]{ratio_f_opt_ph_vs_f_tp_tf_300.ps}
}
\centerline{
\includegraphics[width=5.5cm,angle=270]{ratio_f_x_ph_vs_f_tp_tf_300.ps}
}
\caption{The comparison between the observed in situ flux density at $T_p$ and the
photospheric flux density, at the optical ({\it top}) and X-ray ({\it bottom}) bands,
respectively. The photospheric flux is meaningful only in cases
of $(\tau_e+\tau_{FS}) > 1$ at $T_p$; those cases where $(\tau_e+\tau_{FS}) < 1$
are marked with $f_{\nu}^{ph}= 0$. In cases of $\tau_{FS} > 1$, the emission produced at
$T_p$ is undetectable, and the light curve is dominated by photosphere emission
arriving at a later time; these cases are marked with $f_{\nu}(T_p)= 0$.
When $\tau_e > 1$ and $\tau_{FS} < 1$, the light curve is dominated by either the in situ
emission produced at $T_p$ and diminished by a factor of $\tau_e$ or the photospheric
emission, whichever is larger, so is the observed peak flux.}
\label{fig:ratio-ph-tp}
\end{figure}

\begin{figure}
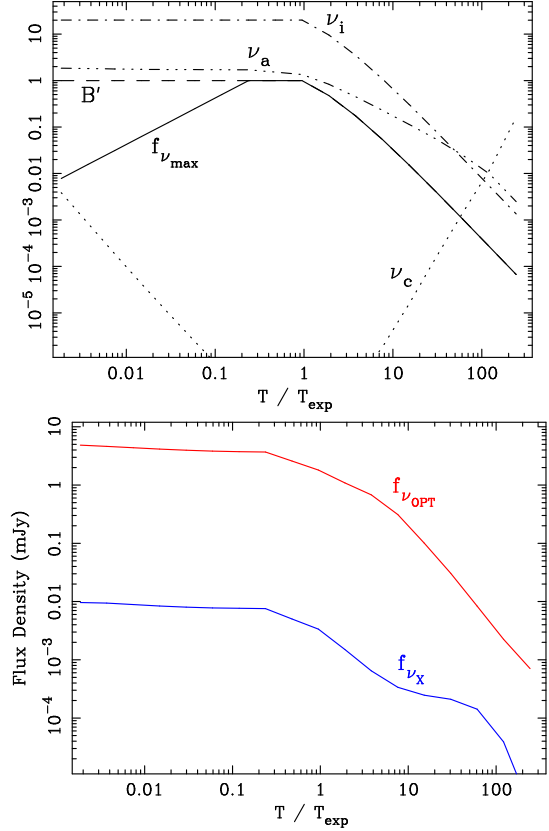

\centerline{
\includegraphics[width=5.5cm,angle=270]{tem_evo_EjEc_0_Gamc_3.ps}
}
\centerline{
\includegraphics[width=5.5cm, angle=270]{lc_EjEc_0_Gamc_3.ps}
}
\caption{{\it Top}: temporal evolutions of $B'$, $\nu_i$, $\nu_c$, $\nu_a$
and $f_{\nu, max}$ during the late jet - cocoon interaction for same model
parameter values as in Fig. \ref{fig:spec}. $\nu_i$, $\nu_c$ and $\nu_a$
are in units of eV, whereas $B'$ and $f_{\nu, max}$ are normalized by their
maximum values, respectively. {\it Bottom}: observed light
curves in optical and X-ray bands. The observer's time $T$ is normalized
by the expansion time scale $T_{exp}= r_i/(2\G_s^2c)$ and
$T= 0$ is when the interaction begins.}
\label{fig:lc-1}
\end{figure}

\begin{figure}
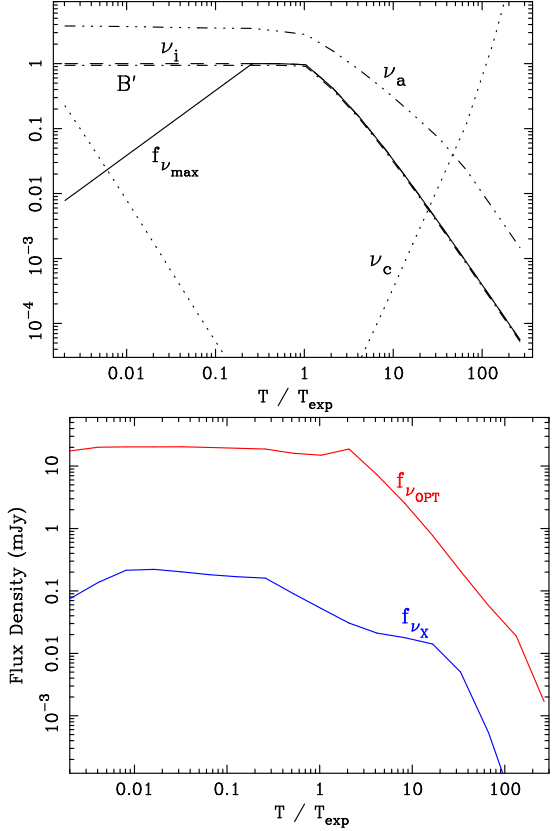

\centerline{
\includegraphics[width=5.5cm,angle=270]{tem_evo_EjEc_1_Gamc_5.ps}
}
\centerline{
\includegraphics[width=5.5cm,angle=270]{lc_EjEc_1_Gamc_5.ps}
}
\caption{{\it Top}: same as Fig. \ref{fig:lc-1} but for
$E_j=10\times E_c=10^{51}$ erg and $\G_c=5$. {\it Bottom}: observed light curves.
Both the optical and X-ray light curves show a flat part at $T < T_{exp}$ because both frequencies
are in the fast cooling regime and the decrease of $\nu_c \propto T^{-2}$ just
compensates for the increase of $f_{\nu, max} \propto T$. The optical peak
at $T>T_{exp}$ is due to the fact that $\nu_c$ increases steeply and $\nu_a$ drops
below $\nu_{opt}$. A rise does not show up in the X-ray light curve at $T>T_{exp}$ because
$\nu_X$ is always above all of $\nu_a$, $\nu_i$ and $\nu_c$.}
\label{fig:lc-2}
\end{figure}


\subsubsection{Optically thick cocoon} \label{sec:opt-thick}

When $r_i$ is small -- either because of a low LF of the cocoon or a small
$t_F$ -- the cocoon could be optically thick to Thomson scattering. For instance,
the optical depth of the entire cocoon at $r_i$ is estimated to be
$\approx 2 \sigma_T E_c/[4\pi r_i^2 \G_c m_p c^2(1-\cos\theta_c)] \sim 0.1 \, E_{c,51} \G_{c,1}^{-5} t_{F,2}^{-2}$.
Figs. \ref{fig:tau_e} and \ref{fig:tau_fs} depict the calculated $\tau_e$ and $\tau_{FS}$
--- the optical depths of the shocked region and the unshocked cocoon,
respectively --- at the expected light curve peak time $T_p$.

When $\tau_e \gg 1$, the photons are subject to numerous scattering
(diffusion) before escaping the plasma. The emergent flux is spread over the
diffusion time scale
\begin{equation}
\D T_d \approx  \frac{r_{ph}}{2\G_s^2 c},
\end{equation}
which is the delay between the actually observed time of a photon and the time it would have been observed in the absence of scattering, where $r_{ph}$ is the photosphere radius.

When $\tau_e \gg 1$, we consider all the photons emitted during the time up to $T_p$ as a photon gas co-expanding with the baryon gas; the expansion of the system is governed by the radiation pressure rather than the gas pressure, and the scattering between photons and electrons is nearly elastic. This treatment is different from the one adopted by Pe'er, Meszaros \& Rees (2006) who consider the case where the gas pressure dominates over the radiation pressure and photons {\it Compton} scatter off thermal electrons.

The equation of state for the photon gas is $(h\nu)^4 \propto V^{-4/3}$, where
$h\nu$ is the characteristic photon energy and $V \propto r^2 \D_c$ is the
volume of the system (in the photon-baryon gas co-expanding phase, the LF of
the system is constant). The width of the system $\D_c$ is $\propto r$
for the thin shell case and is constant for the thick shell case. Since the system
is shock-compressed after the collision, we believe the thin shell case is a more
likely possibility to consider than the thick shell case.
Thus the temporal evolution of the photon energy is $h\nu \propto T^{-1}$, where
$T$ is the observer's time. The shape of the spectrum at the photosphere is
unchanged from that at the time $T_p$.

We first calculate the flux density at $T_p$ neglecting all
optical-thick effects. If $\tau_e > 1$ and $\tau_{FS}< 1$, the promptly observed
flux in situ at $T_p$ is $1/\tau_e$ of that calculated when $\tau_e$ is neglected.
Then we also calculate the emergent flux at the photosphere. The real peak flux
of the light curve is either the observed flux in situ at $T_p$ or the photospheric flux
at later time, whichever is larger. and the light curve would be dominated by that larger
component. In case the observed flux in situ at $T_p$ is stronger than the
photospheric flux, we calculate elaborately the light curve shape following $T_p$ in the
way described in Sec \ref{sec:lc-shape} and numerically estimate its peak time
and pulse width. In case the observed light curve is dominated by the flux at the
photosphere, its peak time is the photospheric time $T_{ph} \sim r_{ph}/(2\G_s^2 c)$
and the pulse width is also $\approx T_{ph}$.
If $\tau_{FS} > 1$, then the unshocked cocoon blocks the light produced in the
shocked region from reaching the observer, thus the emission at the photosphere
is what we actually see only and it will completely dominate the light curve. We show in
Fig. \ref{fig:ratio-ph-tp} a comparison between the observed flux in situ at
$T_p$ and the photospheric flux for the considered model parameter space.


\begin{figure}
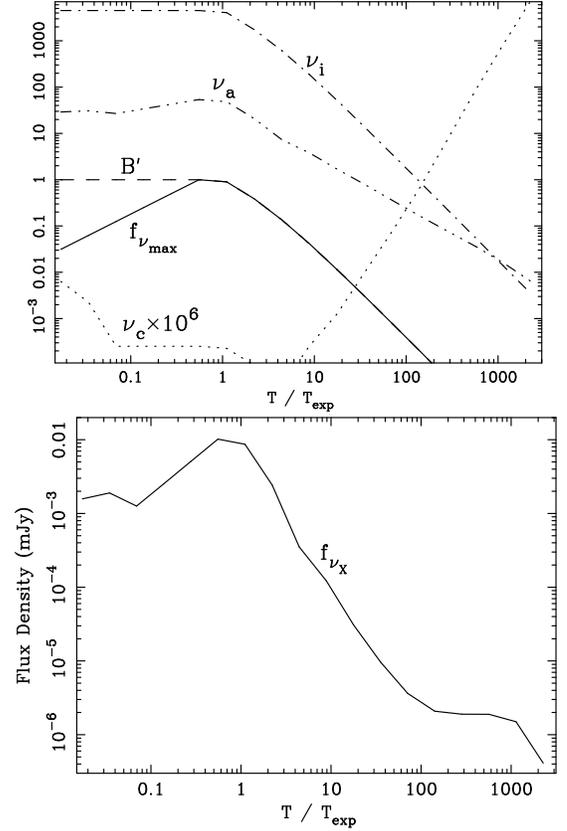

\centerline{
\includegraphics[width=5.5cm,angle=270]{tem_evo_EjEc_2_Gamc_1.12.ps}
}
\centerline{
\includegraphics[width=5.5cm,angle=270]{lc_EjEc_2_Gamc_1.12.ps}
}
\caption{{\it Top}: same as Fig. \ref{fig:lc-1} but for
$E_j=10^2\times E_c=10^{52}$ erg and $\G_c\b_c= 0.5$.
{\it Bottom}: observed light curve in X-rays. The optical light curve is not shown, because for
this set of parameter values the cocoon is extremely optically thick
($\tau_e \gg 1$ and $\tau_{FS}= 0$; see Fig. \ref{fig:tau_e} and
\ref{fig:tau_fs}), and we find the optical light curve is dominated by the photosphere
emission, whose numerical light curve shape has to be calculated differently from that
in the optically thin case. However in the X-ray band, the prompt non-thermal
flux, after diminished by the optical thick effect, is still brighter than the
later photospheric flux. Thus we use the diminished prompt non-thermal flux to
represent the observed X-ray light curve.}
\label{fig:lc-3}
\end{figure}

\begin{figure*}
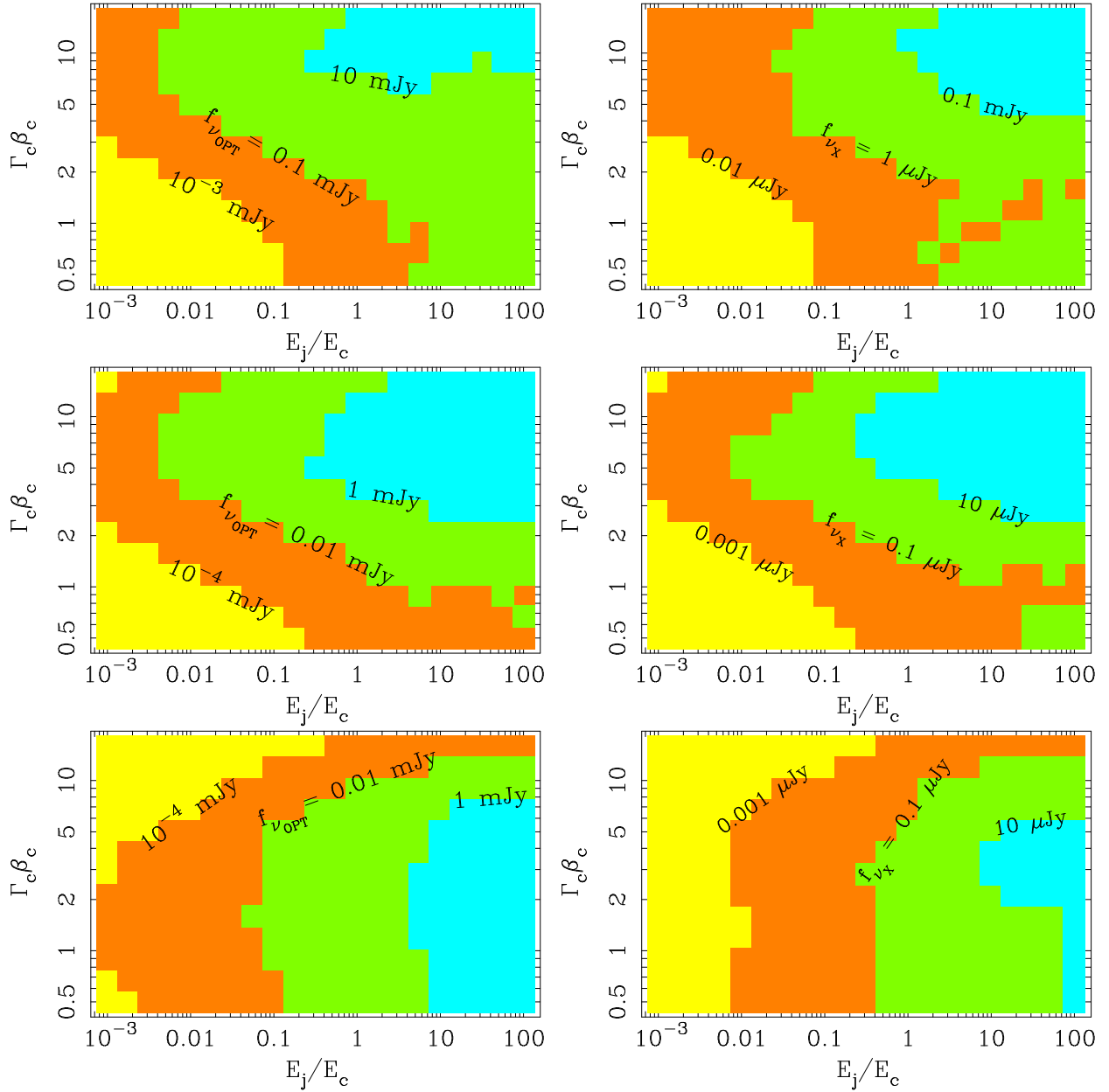

\centerline{
\includegraphics[width=5.5cm,angle=270]{f_opt_tf_100_epsb_-2.ps}
\hspace{0.1cm}
\includegraphics[width=5.5cm, angle=270]{f_x_tf_100_epsb_-2.ps}
}
\centerline{
\includegraphics[width=5.5cm,angle=270]{f_opt_tf_100_epsb_-4.ps}
\hspace{0.1cm}
\includegraphics[width=5.5cm, angle=270]{f_x_tf_100_epsb_-4.ps}
}
\centerline{
\includegraphics[width=5.5cm,angle=270]{f_opt_tf_1e4_epsb_-2.ps}
\hspace{0.1cm}
\includegraphics[width=5.5cm, angle=270]{f_x_tf_1e4_epsb_-2.ps}
}
\caption{The contours of the observed peak flux densities from
the late jet - cocoon interaction. {\it Left panels}: at the optical band ($\nu= 2$ eV).
{\it Right panels}: at the X-ray band ($\nu= 1$ keV).
The flux density values labeled on the contours are the demarcation values for
two neighbouring contour belts. The ratio of the late jet's duration over its
delay $t_{dur}/t_F$ is 0.3 and the redshift is $z= 2$. {\it Top panels}:
for $t_F= 10^2$ s; other parameter values are same as in Fig. \ref{fig:r_RS-r_FS}.
{\it Middle panels}: same as in top panels except for $\eps_B= 10^{-4}$.
{\it Bottom panels}: same as in top panels except for $t_F= 10^4$ s.}
\label{fig:flux-peak}
\end{figure*}

\begin{figure}
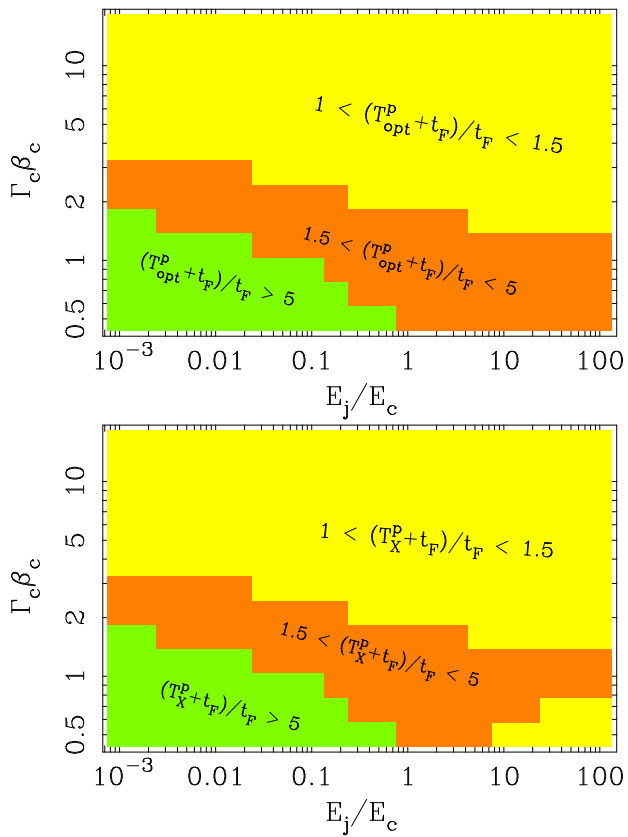

\centerline{
\includegraphics[width=5.5cm,angle=270]{ttp_opt_tf_100.ps}
}
\centerline{
\includegraphics[width=5.5cm,angle=270]{ttp_x_tf_100.ps}
}
\caption{Contours of the light curve peak time for the emission from the
late jet - cocoon interaction, since the burst trigger and normalized by the
delay time of the late jet. {\it Top}: for the optical light curve; {\it Bottom}: for
the X-ray light curve. Model parameter values are $t_{dur}/t_F= 0.3$ and $t_F= 100$ s.
Other parameter values are same as in Fig. \ref{fig:r_RS-r_FS}.}
\label{fig:ttp-tf-300}
\end{figure}

\begin{figure}
\centerline{
\includegraphics[width=5.5cm,angle=270]{ttp_opt_tf_1e4.ps}
}
\centerline{
\includegraphics[width=5.5cm,angle=270]{ttp_x_tf_1e4.ps}
}
\caption{Same as Fig. \ref{fig:ttp-tf-300} except for $t_F= 10^4$ s.}
\label{fig:ttp-tf-3000}
\end{figure}

\begin{figure}
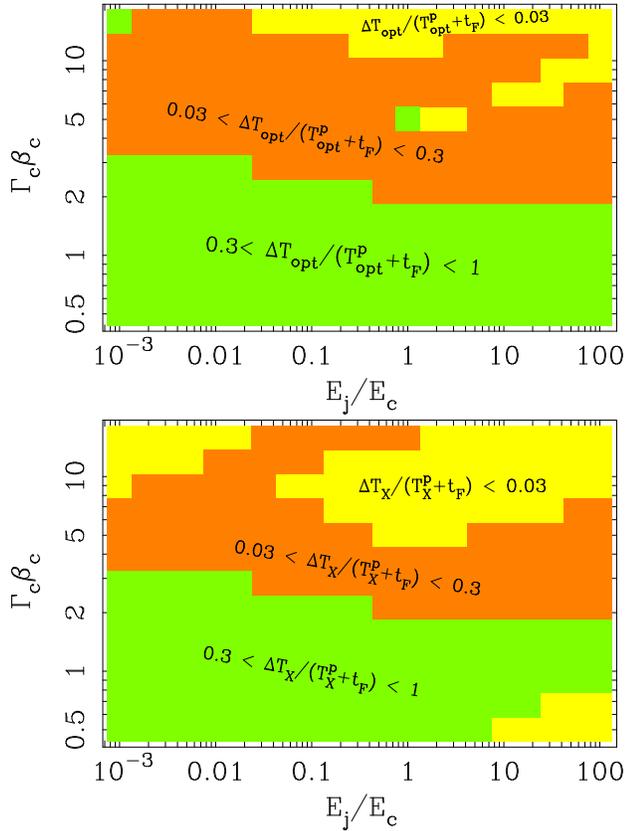

\centerline{
\includegraphics[width=5.5cm,angle=270]{fwhm_opt_tf_100.ps}
}
\centerline{
\includegraphics[width=5.5cm,angle=270]{fwhm_x_tf_100.ps}
}
\caption{Contours of the full width at half maximum (FWHM) of the light curve from
the late jet - cocoon interaction, normalized by the peak time of the light curve since
the burst trigger. {\it Top:} for the optical light curve; {\it Bottom:} for the X-ray light curve.
Model parameter values are $t_{dur}/t_F= 0.3$ and $t_F= 100$ s. Other parameter
values are same as in Fig. \ref{fig:r_RS-r_FS}.}
\label{fig:fwhm-tf-300}
\end{figure}

\begin{figure}
\centerline{
\includegraphics[width=5.5cm,angle=270]{fwhm_opt_tf_1e4.ps}
}
\centerline{
\includegraphics[width=5.5cm,angle=270]{fwhm_x_tf_1e4.ps}
}
\caption{Same as Fig. \ref{fig:fwhm-tf-300} except for $t_F= 10^4$ s.}
\label{fig:fwhm-tf-3000}
\end{figure}

\begin{figure}
\centerline{
\includegraphics[width=8cm]{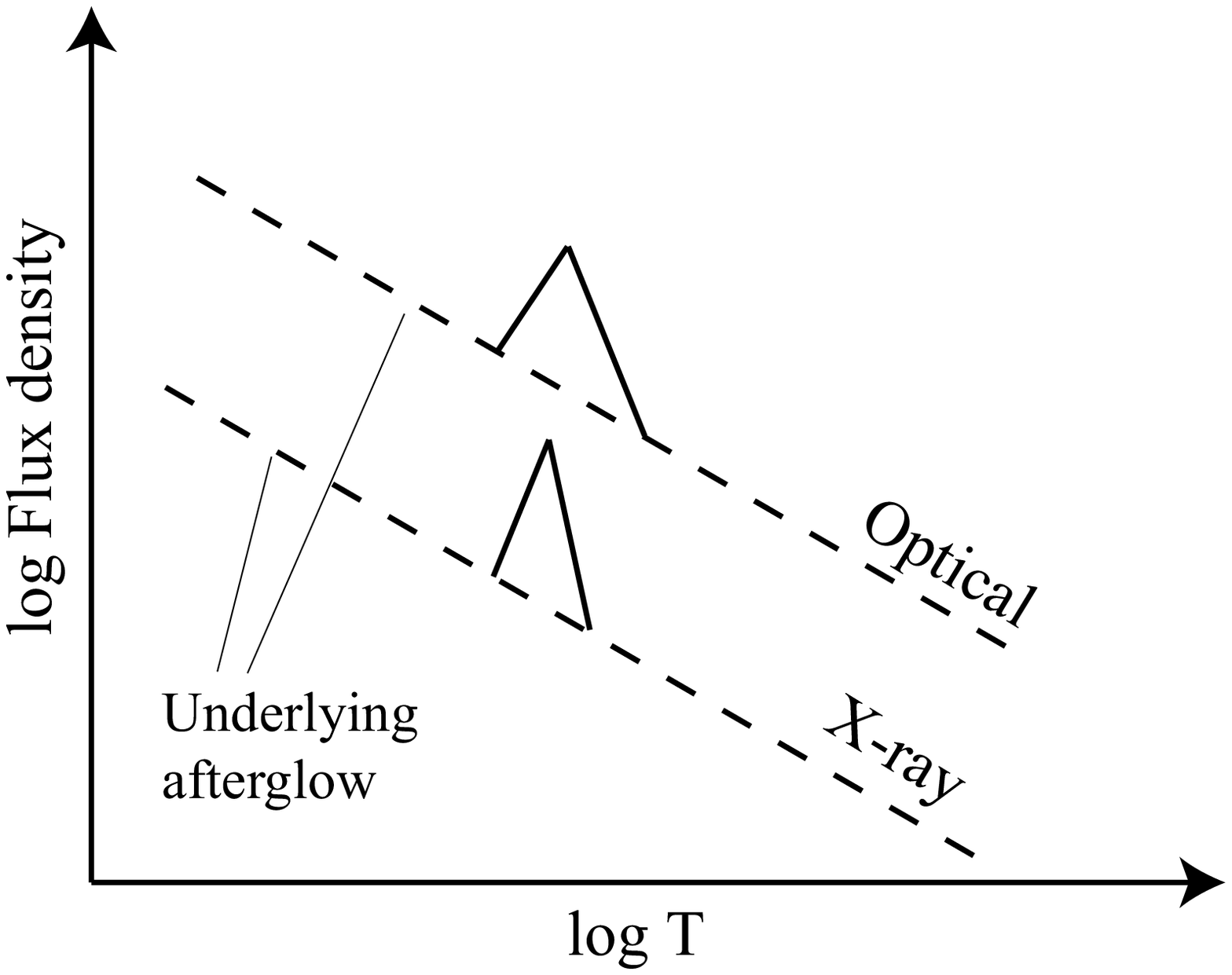}}
\caption{The schematic light curve for the emission from the late jet - cocoon interaction superposed
on the underlying afterglow light curve. Note the zero time in this figure is the
burst trigger time, different from the zero times in Figs. \ref{fig:lc-1} -
\ref{fig:lc-3}.}
\label{fig:lc-illus}
\end{figure}


\subsection{Results}

Figs. \ref{fig:lc-1} - \ref{fig:lc-3} depict the light curves and the temporal evolutions
of emission properties for different values of model parameters. Initially
the light curve remains constant up to the shock crossing time or the expansion time,
whichever is smaller. This is because $\nu_c < \nu_{opt}$, and therefore the
increase of $f_{\nu, max}$ is compensated by the decrease of $\nu_c$. Later,
the X-ray flux decays due to the adiabatic expansion, while the optical flux
continues to rise as long as $\nu_{opt}< \nu_a$. Therefore generally
the optical pulse peaks later and is wider than the X-ray pulse.

The peak flux density distributions are presented in Fig. \ref{fig:flux-peak}. For the range of model parameter values, i.e., $E_j/E_c= 10^{-2} - 10^2$, $\G_c\b_c= 0.5 - 20$, $\eps_B= 10^{-2} - 10^{-4}$, $t_F= 10^2 - 10^4$ s and $E_c= 10^{50}$ erg, the optical peak flux density $f_{\nu_{opt}}$ ranges from $\sim 0.01$ $\mu$Jy to  $\sim 0.1$ Jy, and the X-ray peak flux density is $f_{\nu_X} \approx 0.001$ $\mu$Jy $ - 1$ mJy for $z= 2$. For typical parameter values, i.e., $E_j/E_c= 10^{-1}$, $\G_c\b_c= 3$, $\eps_B= 10^{-2}$ and $t_F= 10^2$ s, the fluxes are $f_{\nu_{opt}} \sim 0.1$ mJy, $f_{\nu_X} \sim 1$ $\mu$Jy.
The ratio of the peak flux densities at optical and X-rays is roughly constant: $f_{\nu_{opt}}/f_{\nu_X} \sim 10^2$, since the optical band is much closer to the spectral peak than the X-rays. The peak flux density is highest for $\G_c \approx 5 - 10$, i.e., when the cocoon is mildly relativistic.

The peak flux densities have very broad ranges. The higher ends of the ranges are high compared to the afterglows, but they correspond to some certain extreme model parameter value (i.e., $E_j/E_c \sim 10^2 - 10^3$) and higher values among model parameter ranges (e.g.,  $\G_c \sim 5 - 10$). The flux values corresponding to typical model parameter values are comparable to those of observed afterglows (see Figs. 17 - 18).

Fig. \ref{fig:flux-peak} also shows the dependence of peak flux densities on $\eps_B$ and $t_F$. When $\eps_B$ varies from $10^{-2}$ to $10^{-4}$, $f_{\nu}$ decreases by a factor of $\sim 10$. This reflects the fact that synchrotron electrons' peak specific radiation power and characteristic frequencies are all linearly dependent on $B'$. When $t_F$ increases from $10^2$ s to $10^4$ s, $f_{\nu}$ drops by a factor of $\sim 10^2$. This is because the interaction radius $r_i \propto t_F$ and therefore both the cocoon and jet densities are smaller at $r_i$ and so is $B'$ for a larger $t_F$. However for $\G_c\b_c < 2$, $f_{\nu}$ increases for a larger $t_F$; this is due to the optical thickness when $t_F$ is small (see Sec. \ref{sec:opt-thick}).

Figs. \ref{fig:ttp-tf-300} and \ref{fig:ttp-tf-3000} show the distribution of peak times of the light curves, according to which the peak time can be approximated by $t_F$, particularly for large values of $t_F$. We measure the full width at half maximum (FWHM) of the light curves to characterize the pulse width, whose distribution is shown in Figs. \ref{fig:fwhm-tf-300} and \ref{fig:fwhm-tf-3000}. The ratio of the pulse width to the peak time, $\D t/t$, has a broad range of 0.01 - 0.5; typically, for lower $\G_c\b_c$ or $E_j/E_c$, pulses are wider. The optical pulses are often wider than X-ray pulses, but the difference is not large.



\section{Observational implication and detection prospects}\label{sec:obs}

\subsection{The Late jet - SNa ejecta interaction}\label{sec:obs-jet-sn}

We find that late jet - SNa ejecta interaction produces a thermal transient due to the breakout of a ``late'' cocoon produced by the late jet crossing the SNa ejecta. Thermal X-ray emission was detected from X-Ray Flash (XRF) 060218 (Campana et al. 2006). But its long lasting, slowly variable light curve suggests that it originated in the shock breakout of a quasi-spherical, mildly relativistic ejecta and the late jet scenario does not apply to this event. For the prevailing X-ray flares detected in GRBs, spectral fit shows no compelling evidence for a thermal component (Falcone et al. 2007).

The non-detection of a thermal X-ray transient in GRBs with flares indicates that
the late jet - SNa ejecta interaction is weak or non- existent. There are three possible reasons for the non-detection:  (i) The cavity in the polar region of the progenitor star created by the main GRB jet is still open, and in that case
the interaction between the late jet and the partially filled cavity is weak resulting in a significantly lower signal than we have estimated in Sec. \ref{sec:the-emi}. The cavity can be kept open, either because the time was too short for the cavity to fill up, or due to a continuous, low-power jet that precedes the late jet.  This low-power jet might also keep the cavity in the cocoon open (see Section \ref{sec:cavity-cocoon}).
(ii) The cocoon that is in front of the ``late'' cocoon and the late jet
could be optical thick, e.g., $\tau_c \sim 0.1 \, E_{c,51} \G_{c,1}^{-5} t_{F,2}^{-2}$ (also see Figs. \ref{fig:tau_e} - \ref{fig:tau_fs}), and block the thermal transient signal. (iii) It is also possible that even though the GRB was associated with the death of a massive star the stellar envelope collapsed producing no supernova at all. In the failed SN explosion the whole stellar envelope would collapse on a free-fall time scale of a couple of hundred seconds (see, e.g., Kumar, Narayan \& Johnson 2008), form a torus and then be accreted. Note that even in this extreme situation, the cocoon associated with the main jet should still exist because the cocoon is created when the head of the prompt GRB jet passes through the stellar envelope, long before the envelope reacts to the core collapse.



\subsection{Late jet - cocoon interaction} \label{sec:obs-jet-coc}

A schematic sketch of light curves due to emission from the late jet - cocoon
interaction, superposed on the underlying external-shock afterglow
component, is illustrated in Fig. \ref{fig:lc-illus}. The expected
emission from late jet - cocoon interaction has the following features:

(1) \textit{Peak flux density}. The peak flux has a broad distribution (see Fig. \ref{fig:flux-peak}).
Except for the cases of a very slow cocoon or a very low jet-to-cocoon
energy ratio, the emission is fairly bright. For instance, the range of
the calculated X-ray peak flux densities corresponds to a flux in the
0.3--10 keV band of $\approx 10^{-14} - 10^{-8}$ erg s$^{-1}$ cm$^{-2}$
(for a spectral index $\b_X = -1.1$) while the Swfit / XRT sensitivity is
$2\times10^{-14}$ erg s$^{-1}$ cm$^{-2}$ in $10^{4}$ s
(Gehrels et al. 2004). Therefore, the X-ray emission from this interaction
is detectable by Swift / XRT for most of the parameter space.

(2) \textit{A small $\D t/t$} ($< 0.5$; see Figs. \ref{fig:fwhm-tf-300} and \ref{fig:fwhm-tf-3000}).

(3) \textit{A non-thermal spectrum.} The emission is mostly non-thermal, except
when the cocoon speed is sub-relativistic and the delay of the late jet is small (e.g., $t_F \lesssim 10^3$ s) thus the thermal photospheric emission might dominate. The X-ray band is in the ``fast cooling'' spectral regime when the emission is non-thermal, and the optical is near the synchrotron self absorption frequency. The flux density ratio of optical and X-ray is roughly $\sim 10^2$ for $p= 2.5$ (this ratio is larger for larger values of $p$). This implies that whenever a X-ray pulse is observed, we expect to see an accompanying increase of  the optical flux.

\begin{figure*}
\centerline{
\includegraphics[width=6.5cm,angle=0]{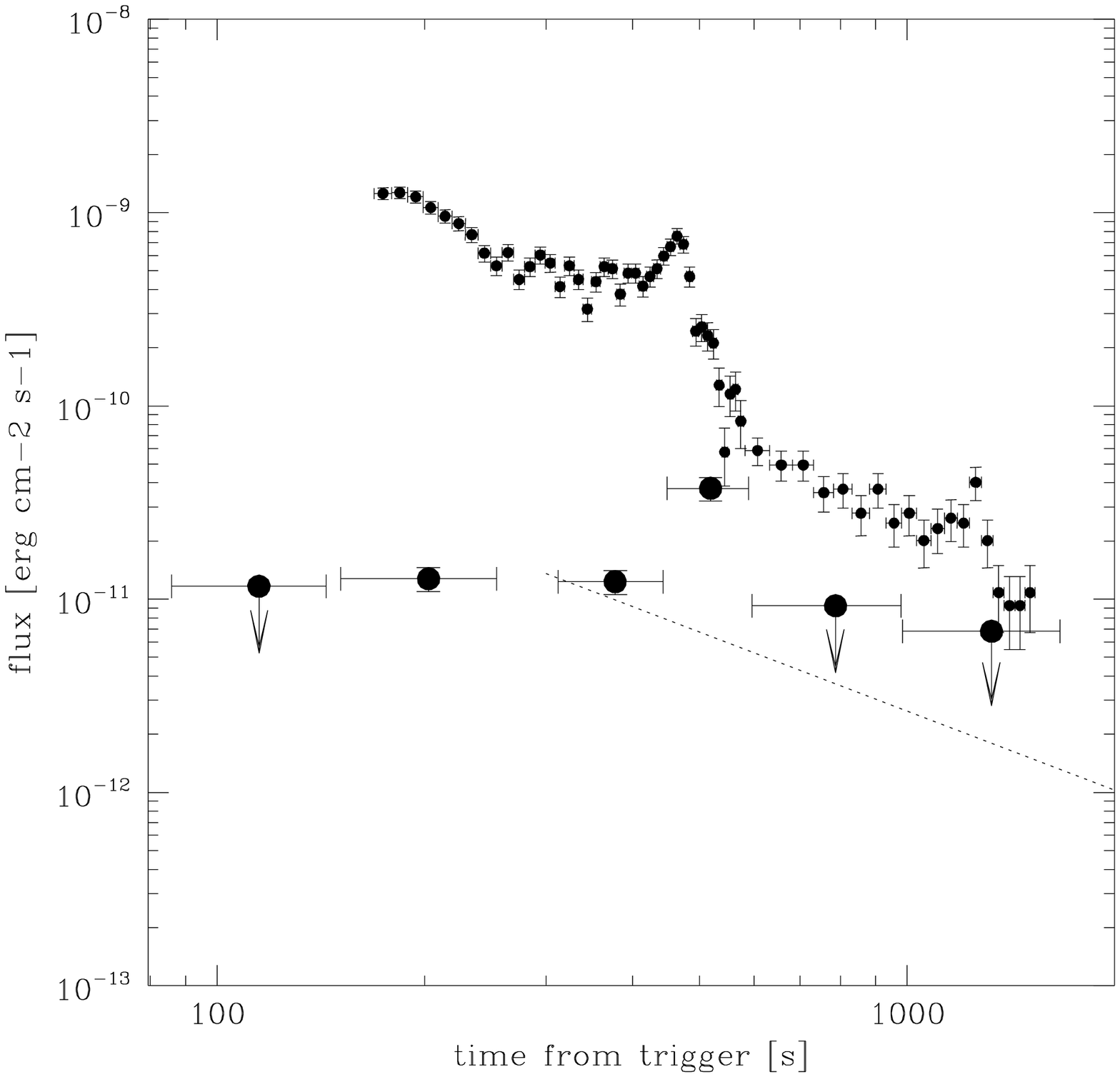}
\hspace{0.1cm}
\includegraphics[width=6.5cm,angle=0]{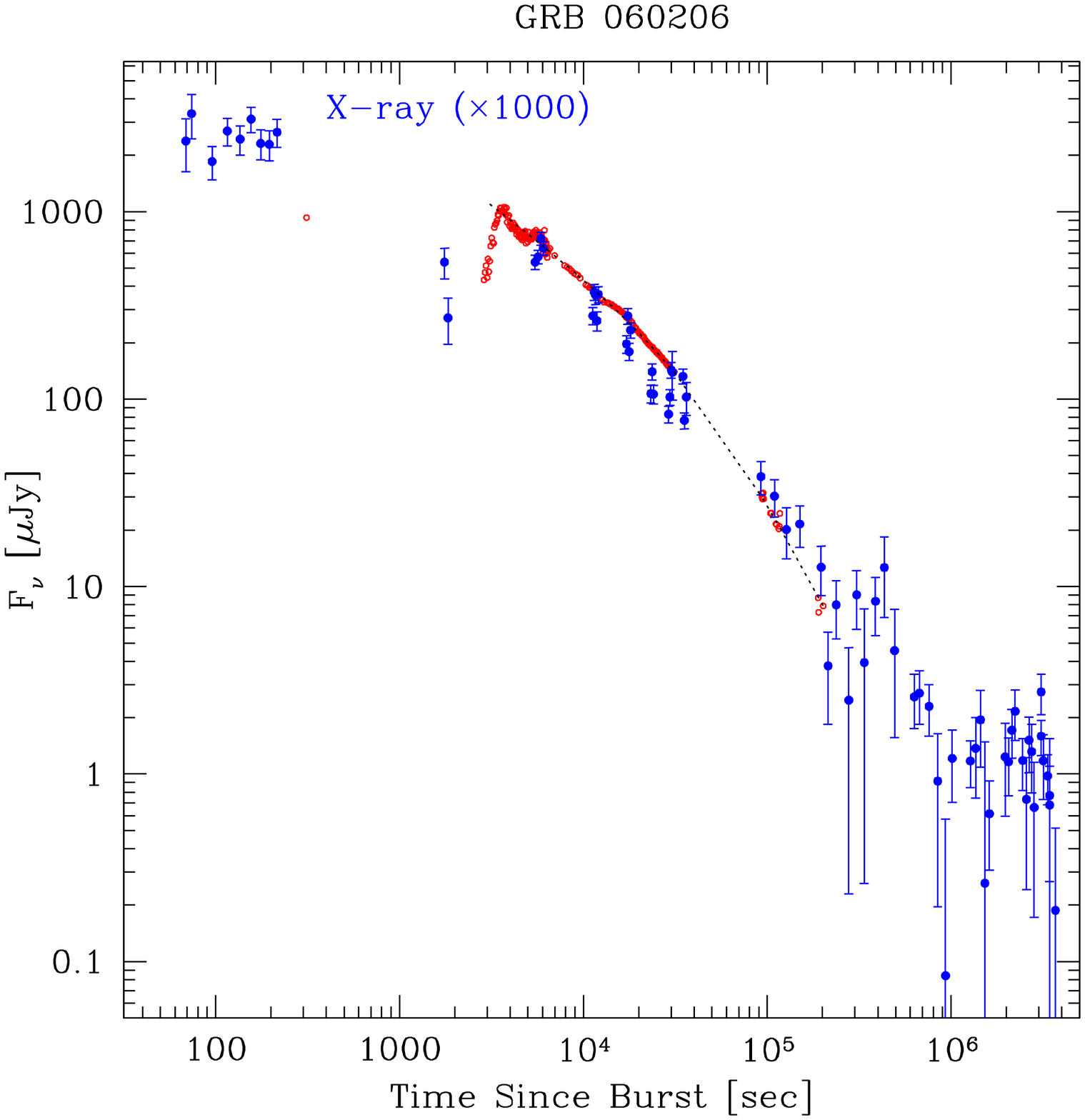}}
\centerline{
\includegraphics[width=6.5cm,angle=0]{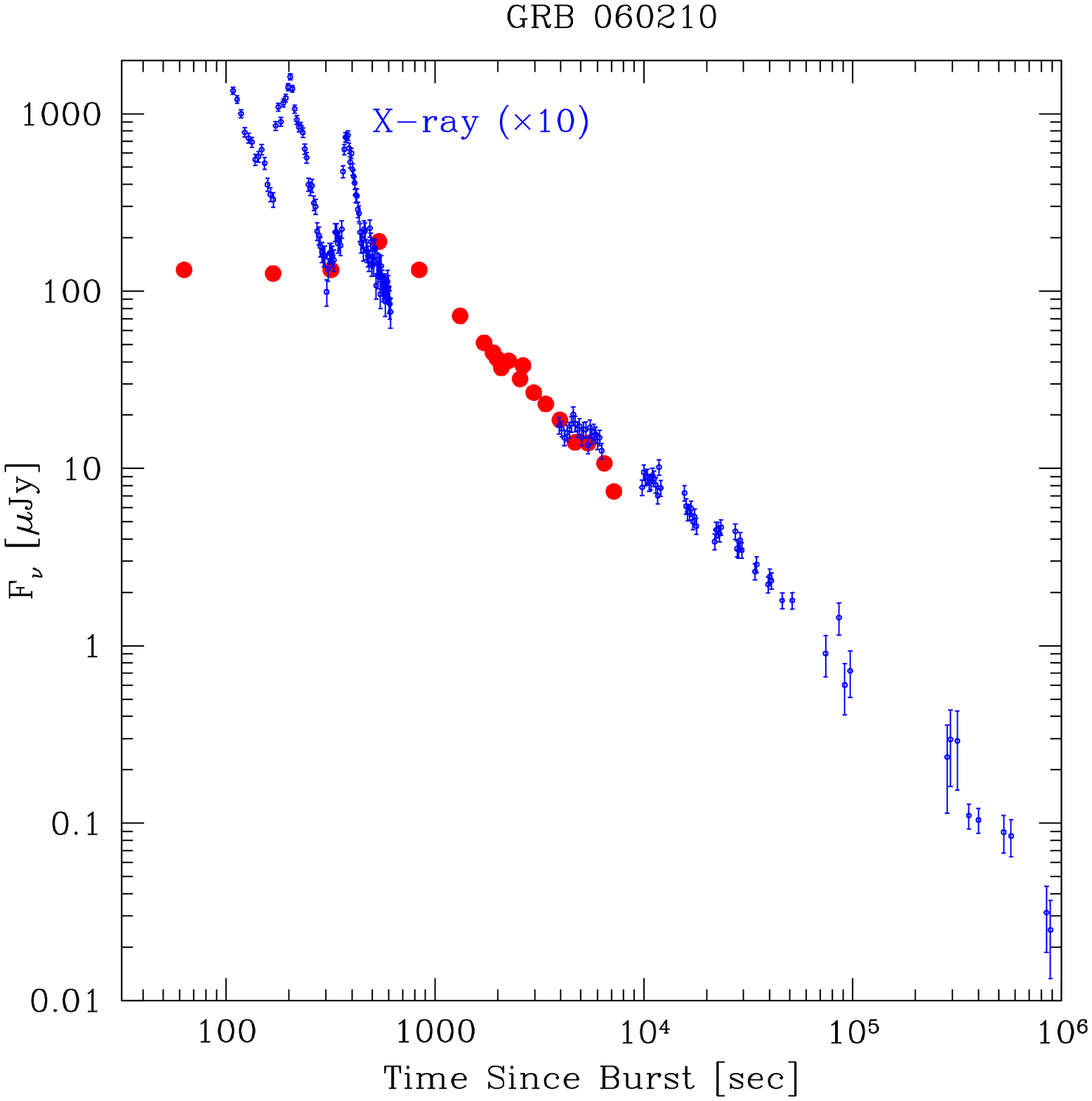}
\hspace{0.1cm}
\includegraphics[width=6.5cm,angle=0]{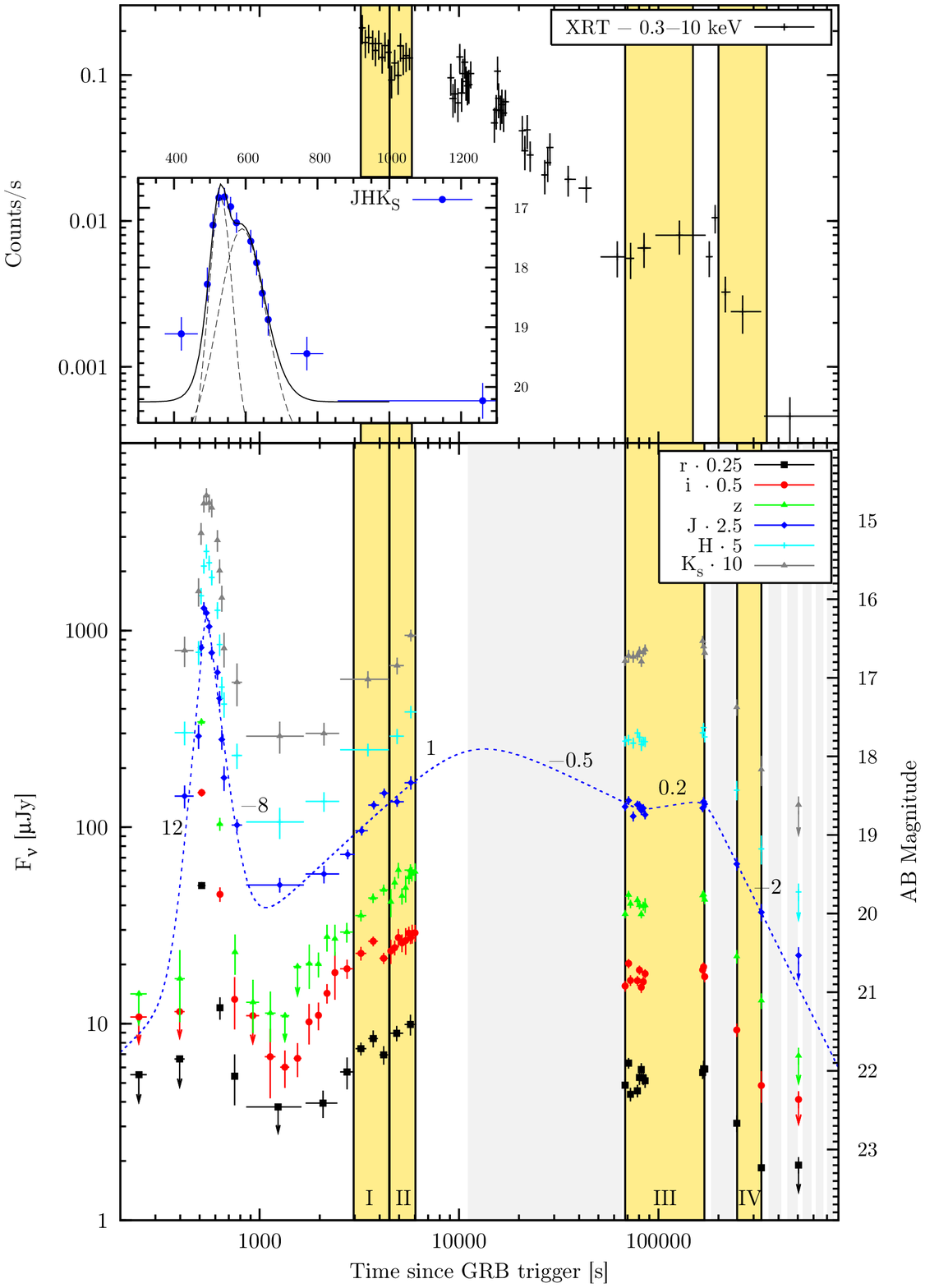}
}
\caption{Four GRBs that show a late flare or rebrightening in optical
afterglow light curve and, in cases where simultaneous X-ray observations are made,
a contemporaneous X-ray flare. {\it Top left}: GRB 050904;
filled circles in red are optical data and filled circles in black are
X-ray data (from Bo\"{e}r et al. 2006). {\it Top right}: GRB 060206, and
{\it bottom left}: GRB 060210; the open and filled circles in red color
(with unnoticeable errors) are the optical data (from Stanek et al. 2007).
{\it Bottom right}: GRB 080129 (from Greiner et al. 2009).} \label{fig:obs-cand}
\end{figure*}

Could some of the late X-ray flares in GRBs originated from the late jet - cocoon interaction? We compare the properties of these flares with the prediction of our calculations in the following.

(1) The observed peak count rate in the XRT band (0.3 - 10 keV) for flares is distributed over the range of 0.1 - 100 counts s$^{-1}$ (Chincarini et al. 2007). Using the empirical instrument conversion factor this translates to $f_{\nu_X}\approx 1 \mu{\rm Jy} - 1 {\rm mJy}$. This range for the observed X-ray flare flux is roughly what is expected for the late jet - cocoon interaction (cf. Fig. \ref{fig:flux-peak}).

(2) The observed value for $\D t/t$ lies in the range of 0.02 to 0.6, with a mean value of XRT band (0.3 - 10 keV) 0.1 (Chincarini et al. 2007). This range is consistent with our calculation (see Figs. \ref{fig:ttp-tf-300} and \ref{fig:ttp-tf-3000}).

(3) It is rare to find an optical flare accompanying a X-ray flare. This is in part due to the fact that very few simultaneous optical observations were made in most cases. Nevertheless, we do find four cases where an optical flare or rebrightening is reported and the X-ray data during the time either is missing or does show a flare. These four cases -- GRB 050904 (shows an optical flare but its X-ray covrage was too sparse to identify a simultaneous flare; Bo\"{e}r et al. 2006), 060206 (Stanek et al. 2007; Wo\'{z}niak et al. 2006), 060210 (Stanek et al. 2007; Curran et al. 2007) and 080129 (Greiner et al. 2009) -- are shown in Fig. \ref{fig:obs-cand}. The last burst shows an early optical flare but without a simultaneous X-ray coverage, and a very late ($2\times10^5$ s since trigger) rebrightening in both optical and X-rays. Among them, the achromatic flarings in GRB 050904, 060210 and 080129 (the very late rebrightening in this burst) are the most likely candidates for a late jet - cocoon interaction event.

On the other hand, there are three cases in which simultaneous optical observations were available but no {\it optical} flare was detected at the time of very strong X-ray flare (the X-ray flux increased by factors $\sim$ 100 in some of these cases), e.g., GRB 060418, 060607A (Molinari et al. 2007) and 060904B (Rykoff et al. 2009) which are shown in Fig. \ref{fig:obs-non-cand}. This shows that not all X-ray flares are due to the late jet - cocoon interaction. However, neither the late jet nor the cocoon is ruled out in these three X-ray flares. The lack of an optical flare indicates that in these cases it is probable that a low-power, continuous jet with a luminosity of $L_{j,low}(t)= 2\times10^{48} t_1^{-2}$ erg s$^{-1}$ (total energy $\sim$ a few $\times 10^{49}$ erg) has kept the cavity open (see  Section \ref{sec:cavity-cocoon}) and the internal shocks between the more powerful late jet and the preceding, slower, low-power jet gave rise to the X-ray flares.



\section{Summary}\label{sec:summary}

Observations of X-ray flares in many GRB afterglows suggest the existence of a late jet from a long-lived central engine of a GRB at  $\sim 10^2$ s but possibly  $10^4 - 10^5$ s after the main GRB event. Adopting the collapsing massive star origin for long-duration GRBs, and assuming that the supernova explosion to be at approximately the same time as the GRB, we have investigated the interactions of this late jet  with the SNa ejecta  and, with a cocoon that was left behind when the main GRB jet traversed the progenitor star.

We find that late jet - SNa ejecta interaction should produce a thermal transient, lasting about $\sim 10$ s and with a peak photon energy at a few keV, that should precede or accompany the flare. The luminosity of this transient is proportional to the kinetic luminosity of the late jet and can be as high as a few $\times 10^{48}$ erg s$^{-1}$. The luminosity is smaller if the polar cavity created by the main GRB jet is only partially filled. This thermal transient is similar to the one associated with the breakout of the main GRB jet. Although it has a lower luminosity, its later occurrence makes it easier to detect. The observation of this signal can provide another evidence for the massive-star origin of GRBs and new information on the GRB - SNa association.

The fact that no such thermal transient was observed so far implies that the late jet - SNa ejecta interaction is suppressed. This could happen if the polar region of the ejecta was evacuated by the main GRB jet and cavity has not sufficiently refilled itself, especially for a not-too-late jet, e.g., $t_F \sim 10^2$ s, or the cavity could be kept open by a continuous, low-power jet. A strong thermal transient signal can also be blocked by the optically thick cocoon, which would be the case if the cocoon is slowly moving. An alternative possibility is of a failed supernova scenario in which the entire stellar envelope collapses in a free fall time of a few $\times 10^2$ s. In this case there is no supernova associated with the GRB and the late jet, if it is late enough,  does not have to cross the stellar envelope.

The late jet interaction with the cocoon would cause a flare or rebrightening, superposed on the afterglow light curves, at both the optical and X-ray bands. This flare would have a pulse-width-to-time ratio $\D t /t < 1$ (the expected distribution of $\D t/t$ is similar to that for X-ray flares). Depending on model parameters, we find for a burst at a redshift $z=2$ that the peak flux density at optical $f_{\nu_{opt}}$ ranges from 0.01 $\mu$Jy to 0.1 Jy ($V$-band apparent magnitude 29 to 11.5) and at X-rays $f_{\nu_X}$ ranges from 0.001 $\mu$Jy to 1 mJy. For typical parameters $f_{\nu_{opt}} \sim 0.1$ mJy ($V$-band magnitude $\sim 19$) and $f_{\nu_X} \sim 1$ $\mu$Jy. Observational identification of this emission would verify the existence of the cocoon produced when the GRB jet traversed the progenitor star, thus it would be another confirmation of the collapsar model for long duration GRBs (e.g., MacFadyen \& Woosley 1999; Ramirez-Ruiz et al. 2002; Matzner 2003; Zhang et al. 2004).

The late jet - cocoon interaction might have already been detected in four GRB afterglows in which simultaneous X-ray and optical flares with $\D t /t \ll 1$ were observed after the prompt emission has died off (see Fig. \ref{fig:obs-cand}). From those candidate events, one can learn about the energetics of late jet and the cocoon by utilizing the emission calculation presented in this paper. Let us consider the flare event in GRB 050904 as an example.  We find the most probable model parameters -- for this burst at $z= 6.3$ and with $t_F= 70$ s -- that produce the observed peak $f_{\nu_{opt}}$ and $f_{\nu_X}$ (data from Bo\"{e}r et al. 2006; Cusumano et al. 2007; Gou, Fox \& M\'{e}sz\'{a}ros 2007) to be $E_c \approx 10^{51}$ erg, $\G_c \approx 20 - 50$, $E_j \approx 10^{52}$ erg and $\G_j \approx 500$. Those high energetics seem consistent with the very luminous nature of both the burst and the flare.

\begin{figure}
\centerline{
\includegraphics[width=6cm,angle=0]{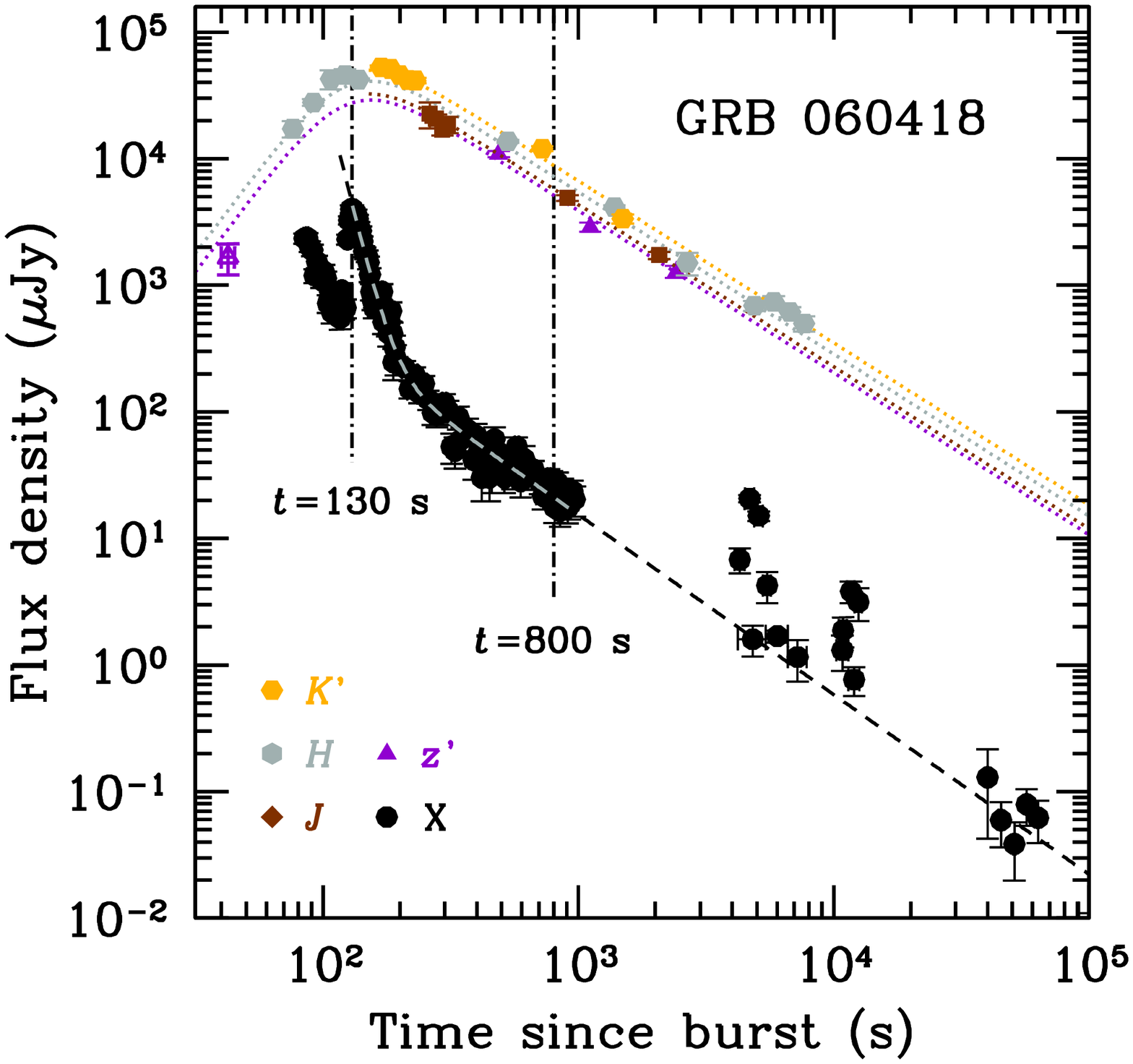}
}
\centerline{
\includegraphics[width=6cm,angle=0]{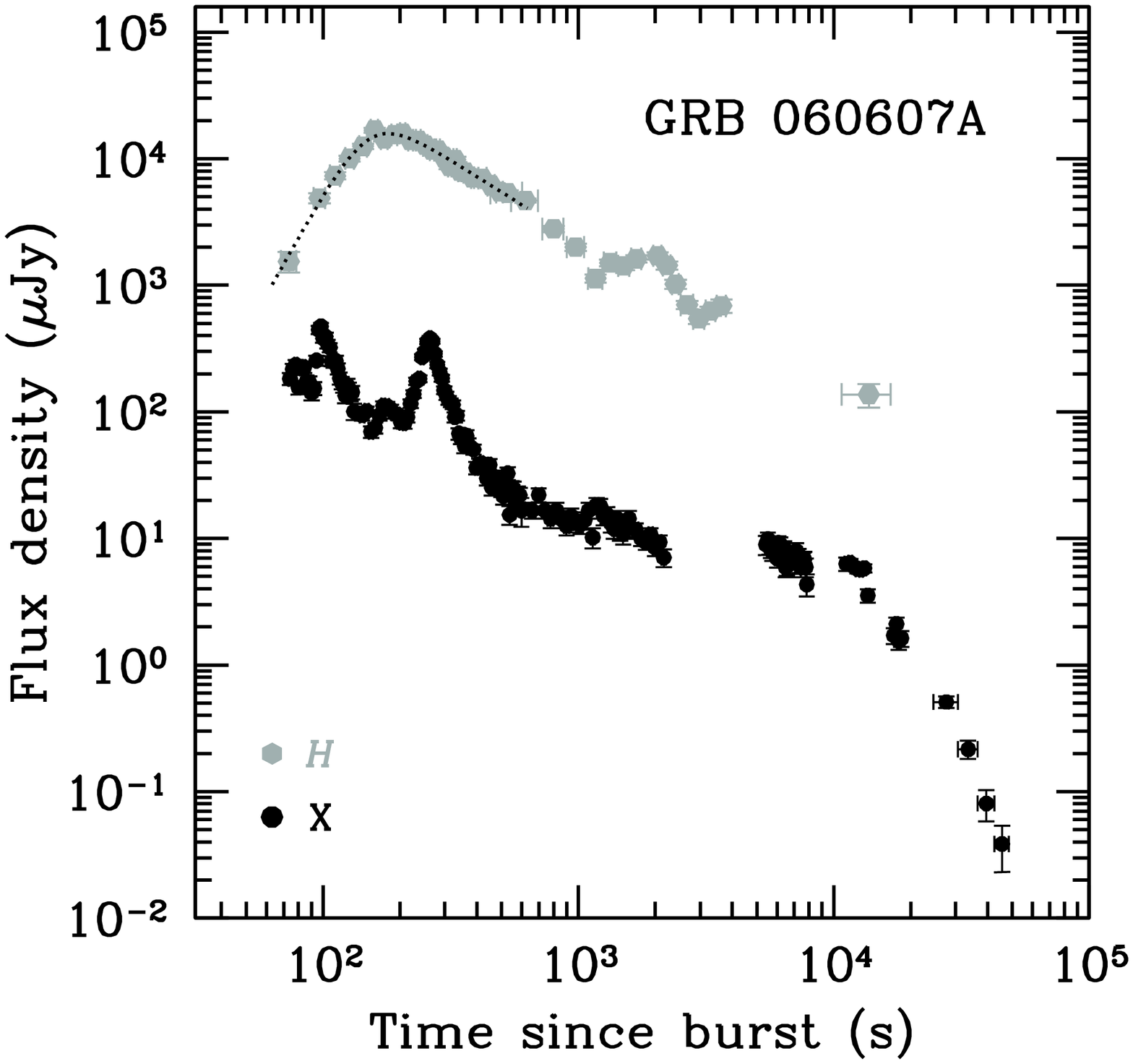}
}
\centerline{
\includegraphics[width=7cm,angle=0]{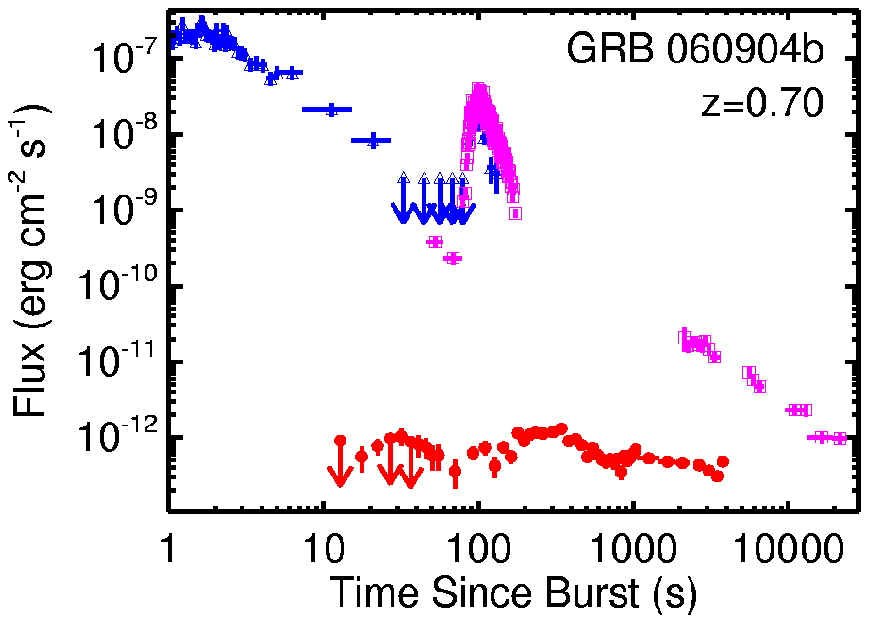}
}
\caption{The three GRBs that show prominent late X-ray flares but without
simultaneous optical flare apparent in the afterglow light curve. {\it Top}:
GRB 060418; {\it Middle}: 060607a (both from Molinari et al. 2007). {\it Bottom}:
GRB 060904b; blue triangles are BAT data extrapolated to X-ray band,
magenta squares are XRT data and red circles are optical data
(from Rykoff et al. 2009).} \label{fig:obs-non-cand}
\end{figure}

There are three cases in which no optical flare was detected at the time of a strong X-ray flare, even though a number of optical telescopes have been observing these bursts at the time of the X-ray flares (Fig. \ref{fig:obs-non-cand}). This shows that not all X-ray flares are due to the late jet - cocoon interaction. However, neither a late jet nor the cocoon can be ruled out in these cases. It is possible that a low-power jet preceding the late jet with a total energy of at least $10^{49}$ erg had kept the cavity in the cocoon open, so that the late jet - cocoon interaction was suppressed.  If correct this implies a low level continuous emission from the central engine at the level of $\sim 10^{47}\, (t/10 \, {\rm s})^{-2}$ erg s$^{-1}$ lasting for $\sim 10^{2}$ s, assuming a radiation efficiency of $\sim 0.1$.


\section*{Acknowledgement}
RS is grateful to Craig Wheeler and Sean Couch for very helpful discussions. This work is supported in part by a NSF grant AST-0909110 (PK), and by an ERC advanced research grant and the ISF Center for High Energy Astrophysics (TP).




\begin{thebibliography}{}
\bibitem{} Barniol Duran R., Kumar P., 2009, MNRAS, 395, 955
\bibitem{} Blandford R. D., McKee C. F., 1976, Physics of Fluids, 19, 1130
\bibitem{} Bloom J. S. et al., 1999, Nature, 401, 453
\bibitem{} Bloom J. S. et al., 2002, ApJ, 572, L45
\bibitem{} Bloom J. S., Kulkarni S. R., Djorgovski S. G., 2002, AJ, 123, 1111
\bibitem{} Bo\"{e}r M., Atteia J. L., Damerdji Y., Gendre B., Klotz A., Stratta G., 2006, ApJ, 638, L71
\bibitem{} B\H{o}ttcher M., Dermer C. D., 2000, ApJ, 532, 281
\bibitem{} Burrows D. N. et al., 2005, Science, 309, 1833
\bibitem{} Burrows D. N. et al., 2007, Phil. Trans. R. Soc. A, 365, 1213
\bibitem{} Campana S. et al., 2006, Nature, 442, 1008
\bibitem{} Castro Cer\'on, J. M. et al., 2006, ApJ, 653, L85
\bibitem{} Chincarini G. et al., 2007, ApJ, 671, 1903
\bibitem{} Christensen L., Hjorth J., Gorosabel J., 2004, A\&A, 425, 913
\bibitem{} Cohen E., Piran T., 1999, ApJ, 518, 346
\bibitem{} Curran P. A. et al., 2007, A\&A, 467, 1049
\bibitem{} Cusumano G. et al., 2007, A\&A, 462, 73
\bibitem{} Della Valle M. et al., 2003, A\&A, 406, L33
\bibitem{} Dermer C. D., 2008, ApJ, 684, 430
\bibitem{} Falcone A. D. et al, 2007, ApJ, 671, 1921
\bibitem{} Fan Y. Z., Wei D. M., 2005, MNRAS, 364, L42
\bibitem{} Fan Y., Piran T., 2006, MNRAS, 369, 197
\bibitem{} Fruchter A. S. et al, 2006, Nature, 441, 463
\bibitem{} Galama T. J. et al., 1998, Nature, 395, 670
\bibitem{} Gehrels N., et al., 2004, ApJ, 611, 1005
\bibitem{} Genet F., Daigne F., Mochkovitch, R., 2007, MNRAS, 381, 732
\bibitem{} Ghisellini G., Celotti A., Ghirlanda G., Firmani C., Nava L., 2007, MNRAS, 382, L72
\bibitem{} Goodman J. 1986, ApJ, 308, L47
\bibitem{} Gou L.-J., Fox D. B., M\'{e}sz\'{a}ros P., 2007, ApJ, 668, 1083
\bibitem{} Granot J., K\"{o}nigl A., Piran T., 2006, MNRAS, 370, 1946
\bibitem{} Granot J., Kumar P., 2006, MNRAS, 366, L13
\bibitem{} Greiner J. et al., 2009, ApJ, 693, 1912
\bibitem{} Hjorth J. et al., 2003, Nature, 423, 847
\bibitem{} Janka H.-Th., Langanke K., Marek A., Mart\'{i}nez-Pinedo G., M\"{u}ller B., 2007, Physics Reports, 442, 38
\bibitem{} Katz J. I., Piran T., 1998, ApJ, 501, 425
\bibitem{} Katz J. I., Piran T., Sari R., 1998, Phy. Rev. Let., 80, 1580
\bibitem{} Kobayashi S., Zhang B., 2007, ApJ, 655, 391
\bibitem{} Kumar P., Granot J., 2003, ApJ, 591, 1075
\bibitem{} Kumar P., Narayan R., Johnson J. L., 2008, MNRAS, 388, 1729
\bibitem{} Landau L. D., Lifshitz E. M., 1959, Fluid Mechanics. Pergamon Press, Addison-Wesley Pub. Co., London, p. 501
\bibitem{} Lazzati D., Perna R., 2007, MNRAS, 375, L46
\bibitem{} Lazzati D., Perna R., Begelman M. C., 2008, MNRAS, 388, L15
\bibitem{} Li Z., Song L. M., 2004, ApJ, 608, L17
\bibitem{} MacFadyen A. I., Woosley S. E., 1999, ApJ, 524, 262
\bibitem{} Malesani D. et al., 2004, ApJ, 609, L5
\bibitem{} Matzner C., 2003, MNRAS, 345, 575
\bibitem{} McMahon E., Kumar P., Piran T., 2006, MNRAS, 366, 575
\bibitem{} M\'{e}sz\'{a}ros P., Laguna P., Rees M. J., 1993, ApJ, 415, 181
\bibitem{} M\'{e}sz\'{a}ros P., 2002, ARA\&A, 40, 137
\bibitem{} Molinari E. et al., 2007, A\&A, 469, L13
\bibitem{} Nakar E., Piran T., Granot J., 2003, New Astronomy, 8, 495
\bibitem{} Nakar E., Granot J., 2007, MNRAS, 380, 1744
\bibitem{} Nousek J. A. et al., 2006, ApJ, 642, 389
\bibitem{} O'Brien P. et al., 2006, ApJ, 647, 1213
\bibitem{} Paczy\'{n}ski B., 1986, ApJ, 308, L43
\bibitem{} Paczy\'{n}ski B., 1998, ApJ, 494, L45
\bibitem{} Panaitescu A., M\'{e}sz\'{a}ros P., Burrows D., Nousek J., Gehrels N., O'Brien P., Willingale R., 2006, MNRAS, 369, 2059
\bibitem{} Pe'er A., M\'{e}sz\'{a}ros P., Rees, M. J., 2006, ApJ, 652, 482
\bibitem{} Piran T., Shemi A., Narayan R., 1993, MNRAS, 263, 861
\bibitem{} Piran T., 1999, Physics Reports, 314, 575
\bibitem{} Piran T., 2005, Rev. Mod. Phys., 76, 1143
\bibitem{} Piro L. et al., 1998, A\&A, 331, L41
\bibitem{} Ramirez-Ruiz E., Celotti A. \& Rees M. J., 2002, MNRAS, 337, 1349
\bibitem{} Rybicki G. B., Lightman A. P., 1979, Radiative Processes in Astrophysics. Wiley-Interscience Press, New York.
\bibitem{} Rykoff E. S. et al., 2009, ApJ, 702, 489
\bibitem{} Sari R., Piran T., Narayan R., 1998, ApJ, 497, L17
\bibitem{} Sari R., Piran T., 1999, ApJ, 520, 641
\bibitem{} Shemi A., Piran T., 1990, ApJ, 365, L55
\bibitem{} Shen R.-F., Zhang B., 2009, MNRAS, 398, 1936
\bibitem{} Stanek K. Z. et al., 2007, ApJ, 654, L21
\bibitem{} Tanaka M. et al., 2009, ApJ, 692, 1131
\bibitem{} Uhm Z. L., Beloborodov A. M., 2007, ApJ, 665, L93
\bibitem{} Woosley S. E., Bloom J. S., 2006, ARA\&A, 44, 507
\bibitem{} Woosely S. E., Heger A., 2006, in Holt S. S., Gehrels N., Nousek J. A., eds, AIP Conf. Proc. Vol. 836, Gamma Ray
Bursts in the Swift Era, American Institute of Physics, Melville, NY, p. 398
\bibitem{} Wo\'{z}niak P. R., Vestrand W. T., Wren J. A., White R. R., Evans S. M., Casperson D., 2006, ApJ, 642, L99
\bibitem{} Wheeler C. J., Yi I., H\"{o}flich P., Wang L., 2000, ApJ, 537, 810
\bibitem{} Wheeler C. J., Meier D. L., Wilson J. R., 2002, ApJ, 568, 807
\bibitem{} Yu Y. W., Dai Z. G., 2007, A\&A, 470, 119
\bibitem{} Yu Y. W., Dai Z. G., 2009, ApJ, 692, 133
\bibitem{} Zhang B., M\'{e}sz\'{a}ros P., 2001, ApJ, 552, L35
\bibitem{} Zhang B., 2006, Advances in Space Research, 40, 1186
\bibitem{} Zhang W., Woosley S. E., Heger A., 2004, ApJ, 608, 365
\end{thebibliography}
\end{document}